\documentclass[12pt]{article}

\usepackage[utf8]{inputenc}
\usepackage{charter}
\usepackage{fullpage}
\usepackage{hyperref}
\usepackage{amssymb}
\usepackage{graphicx}
\usepackage{lipsum}
\usepackage{hyperref}
\usepackage{upgreek}
\hypersetup{hidelinks}
\usepackage[colorinlistoftodos]{todonotes}
\setuptodonotes{inline}
\usepackage[hybrid]{markdown}
\usepackage[total={6.5in,9in},top=1in,headsep=0.1in,headheight=1in]{geometry}
\usepackage[backend=biber,style=numeric,citestyle=numeric-comp,giveninits=true,isbn=false,date=year,sorting=none]{biblatex}
\renewbibmacro{in:}{%
  \ifentrytype{article}{}{\printtext{\bibstring{in}\intitlepunct}}}
\AtEveryBibitem{%
  \clearfield{number}}
\DeclareSourcemap{
  \maps[datatype=bibtex]{
    \map[overwrite]{
      \step[fieldsource=doi, final]
      \step[fieldset=url, null]
      \step[fieldset=eprint, null]
    }  
  }
}
\DeclareSourcemap{
  \maps[datatype=bibtex]{
    \map[overwrite]{
      \step[fieldsource=eprint, final]
      \step[fieldset=url, null]
    }  
  }
}
\newcommand\snowmass{
\begin{center}
  \rule[-0.2in]{\hsize}{0.01in}\\
  \rule{\hsize}{0.01in}\\
  \vskip 0.1in
  Submitted to the Proceedings of the US Community Study\\ 
  on the Future of Particle Physics (Snowmass 2021)\\
 
  \rule{\hsize}{0.01in}\\
  \rule[+0.2in]{\hsize}{0.01in}\\[-2em]
\end{center}
}

\usepackage[firstpage=true]{background}
\backgroundsetup{contents={\parbox{6.5in}{\snowmass}}, scale=1,placement=top,opacity=1,color=black,position={3.25in,1.2in}}

\usepackage{fancyhdr}
\fancypagestyle{plain}{%
  \fancyhf{}%
  \fancyhead[C]{}
  \fancyfoot[C]{\thepage}
}

\fancypagestyle{empty}{%
  \fancyhf{}%
  \fancyhead[C]{{\it Snowmass2021 Cosmic Frontier Dark Matter Direct Detection to the Neutrino Fog}}
  \fancyfoot[C]{\thepage}
}
\title{Snowmass2021 Cosmic Frontier \\ Dark Matter Direct Detection to the Neutrino Fog}
\date{}

\usepackage{authblk}

\author[22,23]{D.S.~Akerib}

\author[21]{P.~B.~Cushman \thanks{corresponding author: cushman@umn.edu}}

\author[5,6]{C.~E.~Dahl}

\author[14,15]{R.~Ebadi}

\author[22,23]{A.~Fan}

\author[24]{R.~J.~Gaitskell \thanks{corresponding author: richard\_gaitskell@brown.edu}}
\author[17]{C.~Galbiati \thanks{corresponding author: galbiati@princeton.edu}}
\author[18]{G.~K.~Giovanetti}
\author[8]{Graciela~B.~Gelmini}
\author[19,20]{L.~Grandi}

\author[7]{S.~J.~Haselschwardt}

\author[1]{C.~M.~Jackson}

\author[2]{R.F.~Lang}
\author[1]{B.~Loer \thanks{corresponding author: ben.loer@pnnl.gov}}
\author[13]{D.~Loomba}

\author[16]{M.~C.~Marshall}
\author[13]{A.~F.~Mills}

\author[11,12]{C.~A.~J.~O'Hare}

\author[17]{C.~Savarese}
\author[10]{J.~Schueler}
\author[4]{M.~Szydagis}

\author[9]{Volodymyr Takhistov}
\author[3]{Tim~M.~P.~Tait}
\author[3]{Y.-D.~Tsai}

\author[10]{S.~E.~Vahsen}

\author[14,15]{R.~L.~Walsworth}
\author[17]{S.~Westerdale}

\affil[1]{Pacific Northwest National Laboratory, Richland, WA, USA}
\affil[2]{Department of Physics and Astronomy, Purdue University, West Lafayette, IN 47907, USA}
\affil[3]{Department of Physics and Astronomy, University of California, Irvine, CA 92697, USA}
\affil[4]{The University at Albany, The State University of New York, Albany, NY 12222-0100, USA}
\affil[5]{Department of Physics and Astronomy, Northwestern University, Evanston, Illinois 60208, USA}
\affil[6]{Fermi National Accelerator Laboratory, Batavia, Illinois 60510, USA}
\affil[7]{Lawrence Berkeley National Laboratory, 1 Cyclotron Road, Berkeley, CA 94720, USA}
\affil[8]{Department of Physics and Astronomy, University of California, Los Angeles, Los Angeles, California, 90095-1547, USA}
\affil[9]{Kavli Institute for the Physics and Mathematics of the Universe (WPI), UTIAS, The University of Tokyo, Kashiwa, Chiba 277-8583, Japan}
\affil[10]{Department of Physics and Astronomy, University of Hawai`i, Honolulu, HI 96822, USA}

\affil[11]{The University of Sydney, School of Physics, NSW 2006, Australia}
\affil[12]{ARC Centre of Excellence for Dark Matter Particle Physics, Sydney, Australia}
\affil[13]{Department of Physics and Astronomy, University of New Mexico, NM 87131, USA}
\affil[14]{Department of Physics, University of Maryland, College Park, Maryland 20742, USA}
\affil[15]{Quantum Technology Center, University of Maryland, College Park, Maryland 20742, USA}
\affil[16]{Time and Frequency Division, NIST, Boulder, Colorado 80305, USA}
\affil[17]{Department of Physics, Princeton University, Princeton, NJ 08544, USA}
\affil[18]{Physics Department, Williams College, Williamstown, MA 01267, USA}
\affil[19]{Department of Physics, The University of Chicago, Chicago, Illinois 60637, USA}
\affil[20]{Kavli Institute for Cosmological Physics, Chicago, Illinois 60637, USA}
\affil[21]{School of Physics and Astronomy, University of Minnesota, Minneapolis, MN 55455, USA}
\affil[22]{SLAC National Accelerator Laboratory, Menlo Park, CA 94025-7015, USA}
\affil[23]{Kavli Institute for Particle Astrophysics and Cosmology, Stanford University, Stanford, CA 94305-4085, USA}
\affil[24]{Department of Physics, Brown University, Providence, RI 02912, USA}

\newcommand{\paratitle}[1]{\paragraph{#1}\mbox{}\\}

\begin{document}

\maketitle
\begin{abstract}
We present a summary of future prospects for direct detection of dark matter within the GeV/$c^2$ to TeV/$c^2$ mass range. This is paired with a new definition of the neutrino fog in order to better quantify the rate of diminishing returns on sensitivity due to irreducible neutrino backgrounds. A survey of dark matter candidates predicted to fall within this mass range demonstrates that fully testing multiple well-motivated theories will require expanding the currently-funded generation of experiments down to and past the neutrino fog. We end with the status and plans for next-generation experiments and novel R\&D concepts which will get us there. 
\end{abstract}
\newpage
\section*{Executive Summary}
\label{sec:execsummary}
\begin{itemize}
    \item There are multiple well-motivated dark matter candidates remaining in the ``traditional'' $\sim$GeV-scale mass range. The currently-funded suite of searches in this mass range (``Generation 2'' experiments) will not have sensitivity to fully test the majority of these candidates. 
    \item A ``Generation 3'' suite of experiments with an order of magnitude larger exposure would be able to fully test some candidates. Such a suite should include searches for spin-dependent interactions, which can uniquely test some models not probed by spin-0 targets. 
    \item Despite the maturity of the field, novel technologies should not be neglected. Current R\&D on new techniques will improve established  detector performance and provide new methods to mitigate backgrounds and probe complementary parameter space near the neutrino fog. 
    \item Substantial well-motivated parameter space will yet remain if dark matter signals are not observed by the G3 experiments. Irreducible neutrino backgrounds will cause substantially diminished returns on further increases in exposure. However, if the uncertainties in the neutrino fluxes are reduced, further increases may become feasible. Moreover, light, spin-dependent targets such as fluorine have substantially lower neutrino backgrounds and can therefore scale to larger masses even with current neutrino flux uncertainties. 
    \item Directional detectors are one possible way to reject neutrino backgrounds and thereby reach beyond the neutrino-limited point of current technology. Such detectors will require substantial R\&D investment to reach the size and level of background control required to explore this parameter space. Gas TPCs with micropattern gaseous detector (MPGD) readout should be advanced to the 10 m$^3$ scale. 
\end{itemize}

\tableofcontents

\section{Introduction}
\label{sec:intro}
In this paper, we consider the current and future state of direct experimental searches for interactions of nuclei with dark matter masses approximately in the GeV to TeV range. This mass range, roughly corresponding to what has traditionally been labeled Weakly Interacting Massive Particle (WIMP) dark matter,  has historically received the most focus. Consequently, there is a broad competitive landscape of different targets and mature technologies. 
Complementary papers cover lighter~\cite{SnowmassCF1WP2} and heavier~\cite{SnowmassCF1WP8} direct detection searches as well as indirect searches via astronomical observation~\cite{SnowmassCF1WP5}. 

For the most part, sensitivity in this mass range does not depend strongly on experiment threshold. Rather, improvements in sensitivity are driven primarily by increasing exposure (i.e., target mass) and/or reducing backgrounds. Even for mature technologies that have already fielded multiple generations of detectors, scaling up the detector size presents numerous technical challenges. The most significant shared challenge is that backgrounds passing all fiducial and analysis cuts, both from radioactivity and instrument noise, must decrease proportionally with detector mass. Similarly calibration of detector response functions must improve with every successive generation. More details of these common concerns are presented in \cite{SnowmassCF1WP3}. Other challenges are unique to a given detector technology, such as affordably maintaining high signal collection efficiency while covering a larger area, further from the fiducial volume. To ensure success in further searches, R\&D investments both addressing how to scale up existing technologies as well as examining totally new technologies are warranted. 

If not limited by other factors, dark matter detectors will be limited by irreducible backgrounds from neutrinos~\cite{monroe2007, strigari2009, billard2014b}. Originally dubbed the ``neutrino floor,''  the community now promotes the term ``neutrino fog,'' to better indicate that, rather than a hard limit on direct detection sensitivity, the neutrino background imposes a gradual penalty that can be overcome, at least to some extent. In Section~\ref{sec:neutrinofloor} we quantify the effect of neutrino backgrounds on detector sensitivity versus exposure. 

In Section~\ref{sec:theory} we provide a survey of theoretically-motivated dark matter candidates in this mass range consistent with experimental evidence from dark matter searches, collider experiments, and astronomical observatories. As we shall show, this mass range is well-covered by predictive theories, down to and well into the neutrino fog. Very few of these theories will be fully tested by the upcoming generation of funded experiments. As described in Section~\ref{sec:currentexperiments}, the \textit{subsequent} generation of current experiments (``generation 3'') is expected to reach deep enough into the neutrino fog to incur significant diminishing returns on further growth.  Further exploration of this theoretically-motivated parameter space will require either extreme scaling of detector size or developing new technology less sensitive to neutrino backgrounds, as discussed in Section~\ref{sec:beyondnufog}.

\begin{figure}
    \centering
    \includegraphics[width=\textwidth]{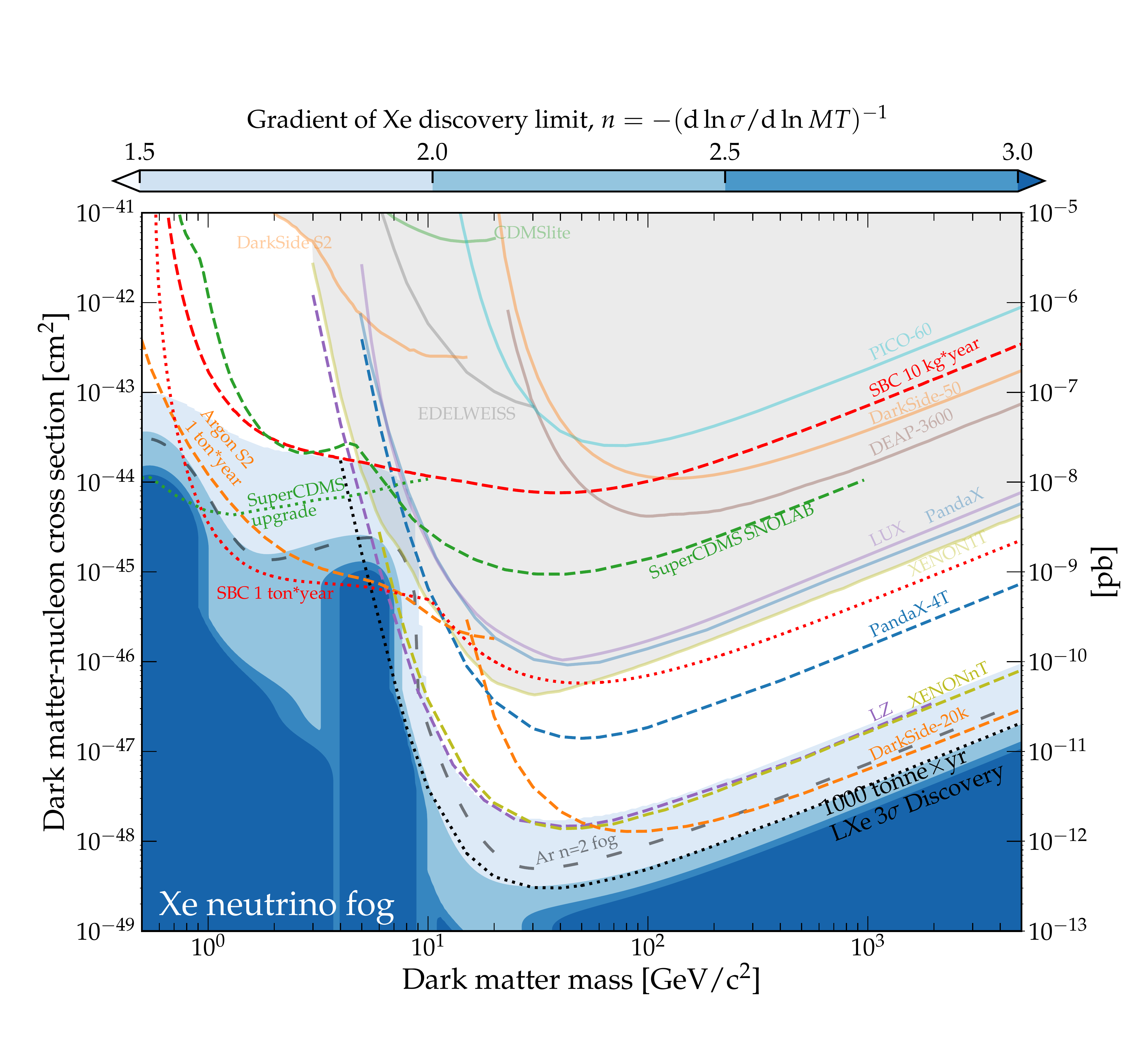}
    \caption{Combined Spin-independent dark-matter nucleon scattering cross section space. Currently-excluded space is shaded gray~\cite{pandax-iicollaboration2017, picocollaboration2017, darksidecollaboration2018a, cresstcollaboration2019,hehn2016a,darksidecollaboration2018,luxcollaboration2017,xenoncollaboration2021,deapcollaboration2019,supercdmscollaboration2019a} (data points taken from~\cite{OHare:2021utq}). Dashed lines represent projected 90\% confidence level exclusion sensitivity of new experiments. Because not all experiments used the same methodology in estimating limits (e.g., single-sided upper likelihood vs two-sided), exact sensitivities may not be directly comparable. The neutrino fog for a xenon target is presented in the blue contour map as described in Section~\ref{sec:neutrinofloor}. At contour $n$, obtaining a 10$\times$ lower cross section sensitivity requires an increase in exposure of at least $10^n$. The $n=2$ fog contour for argon is also shown in the black wide-dashed line. LZ: 15 ton-year, one-sided upper limit. XENONnT: 20 ton-year with two-sided interval. PandaX-4T: 5.6 ton-year. DarkSide-20K: 200 ton-year. SuperCDMS: combined result of detector types; SuperCDMS upgrade refers to scenario C in Ref.~\cite{SuperCDMS_SNOWMASS22}. 
    \label{fig:limitplot_SI}}
\end{figure}

Figure~\ref{fig:limitplot_SI} summarizes the current state of spin-independent dark matter cross section limits as well as the projected sensitivity of future experiments. For technologies so far in use or planned for the near future, the spin-dependent case is dominated by xenon for dark matter-neutron interactions and fluorine for dark matter-proton interactions. Fluorine additionally is several orders of magnitude less sensitive to coherent neutrino interactions than xenon and thus can reach significantly lower spin-dependent cross sections. 

\section{The neutrino fog}
\label{sec:neutrinofloor}

\begin{figure}
    \centering
    \includegraphics[width=0.75\textwidth]{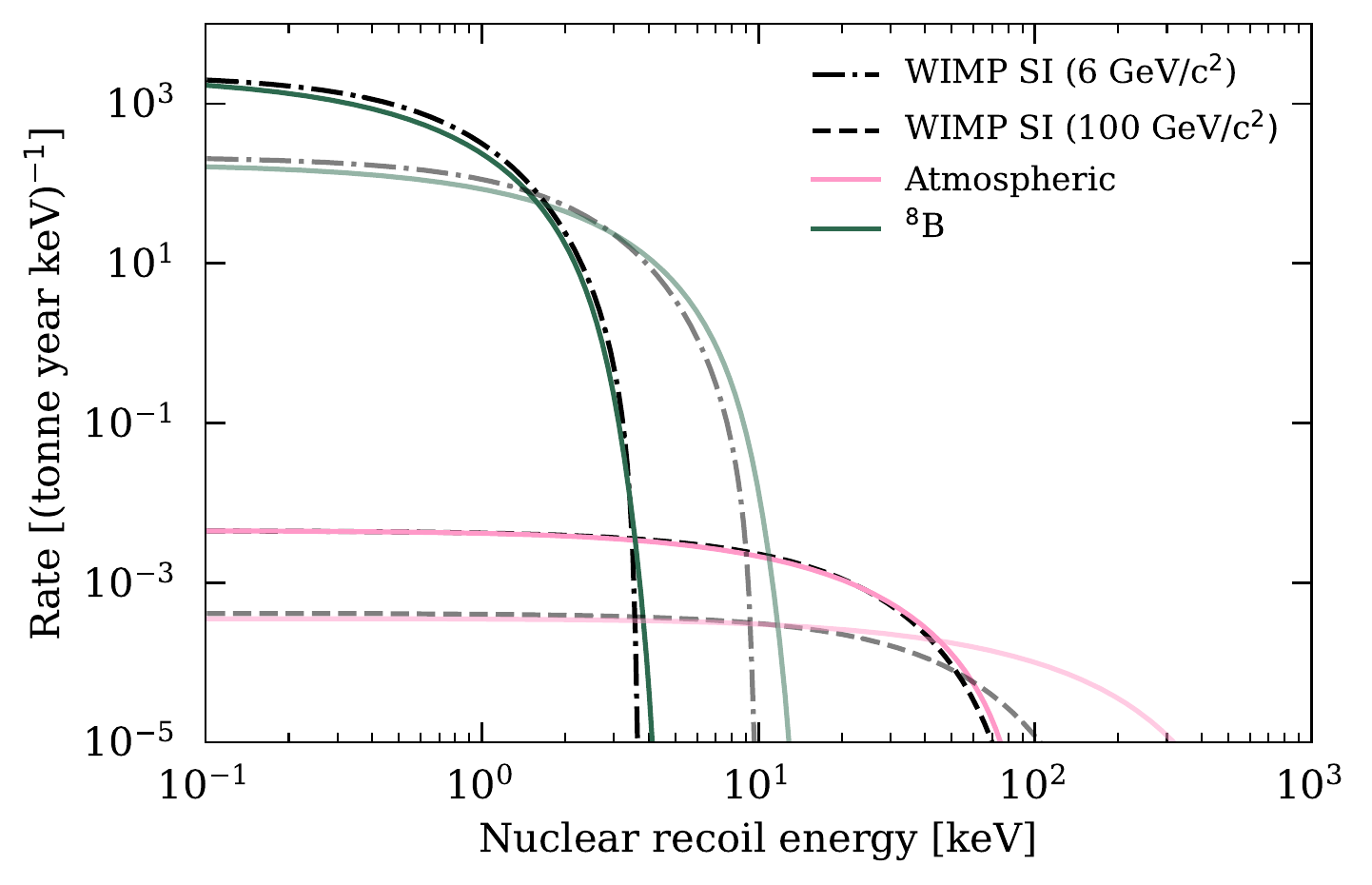}
    \caption{Spectrum of solar-neutrino nuclear recoils scattering on Xe (darker color) and Ar (lighter color). The recoil spectrum for a 6 GeV/$c^2$ (dashed line) and a 100 GeV/$c^2$ (dotted-dashed line) are also given for reference.}
    \label{fig:coherent_solarnu_rates}
\end{figure}

As direct dark matter searches accumulate ever larger exposures in pursuit of testing smaller interaction cross sections, the expected background from the coherent elastic neutrino-nucleus scattering (CE$\nu$NS) of astrophysical neutrinos increases. For dark matter masses $\gtrsim10$~GeV/$c^2$, the relevant neutrino fluxes are those from $^8$B solar neutrinos, atmospheric neutrinos, and the diffuse supernova neutrino background (DSNB). The CE$\nu$NS event rate from each of these is subject to a systematic uncertainty which manifests from the uncertainty on the flux of each neutrino species. In addition, the nuclear recoil energy spectra of these events closely resembles that of the sought after dark matter signal -- this is the case for a spin-independent (SI) dark matter interaction as well as for other couplings/interactions. An example is given in Fig.~\ref{fig:coherent_solarnu_rates}, which shows the SI nuclear recoil rate of a couple of WIMP masses on xenon and argon alongside those of the dominant neutrino backgrounds. 

The effect of these backgrounds -- particularly because of their associated systematic uncertainties and spectral shapes --  will be to reduce the dark matter sensitivity achievable as the experimental exposure grows. This fact led to the creation of the colloquially known ``neutrino floor": a boundary in the cross section versus dark matter matter mass plane below which a dark matter discovery becomes extremely challenging without further constraints on the neutrino backgrounds. The precise location of this boundary has evolved over the last decade~\cite{Billard:2013qya,Gelmini:2018ogy,OHare:2020lva,OHare:2021utq}, however more recent studies~\cite{OHare:2020lva,OHare:2021utq} have emphasized a crucial point: that astrophysical neutrino backgrounds do not impose a hard limit on physics reach; rather, the effect is more gradual than a single boundary depicts. To reinforce this concept within the community, the term ``neutrino fog" has instead been used to describe the region of dark matter cross sections where neutrino backgrounds begin to inhibit the progress of direct detection searches.

\subsection*{Defining the neutrino fog}

We choose to adopt the methodology of O'Hare~\cite{OHare:2021utq} in defining the neutrino fog region, with astrophysicsl parameters updated to match the recommendation in Ref.~\cite{recommendation_dark_matter}. Specifically, the quantity of interest is the index $n$, defined as the gradient of a hypothetical experiment's median cross section for $3\sigma$ discovery with respect to the exposure:

\begin{equation}
    n = -\bigg( \frac{d \log{\sigma}}{d \log{MT}} \bigg)^{-1}
\end{equation}

This index quantifies the diminishing return-on-investment in increasing exposure when limited by neutrinos. Specifically, if an experiment has achieved some cross-section sensitivity, further reducing the sensitivity by a factor of $x$ requires increasing the exposure by at least $x^n$. E.g., at $n=2$, reducing the cross section reach by a further factor of 10 requires increasing the exposure by a factor of at least 100. 

\begin{figure}
    \centering
    \includegraphics[width=0.8\textwidth]{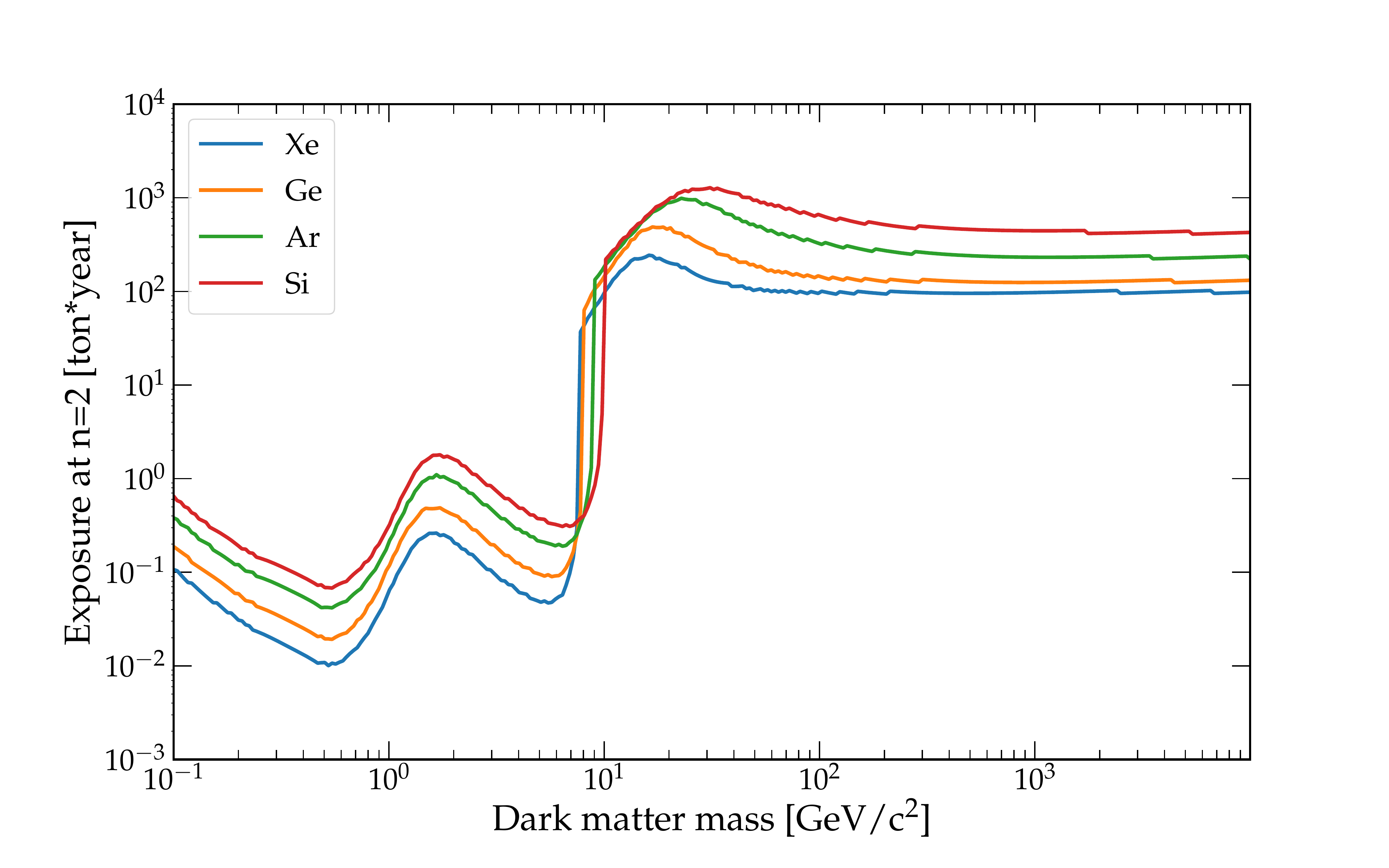}
    \caption{Exposure in ton-years required to reach the $n=2$ (systematic-limited) spin-independent neutrino fog level as a function of dark matter mass for various targets. 
    \label{fig:exposure_at_floor_SI}}
\end{figure}

In order to illustrate the ever-diminishing returns due to the neutrino fog, in Figure~\ref{fig:limitplot_SI}, we plot contours of the index. In order to simplify the contours, the maximum index is projected down.  If the neutrino fog must be represented by a single line, we choose $n=2$, as this marks the transition from statistically- to systematically-limited. Figure~\ref{fig:exposure_at_floor_SI} illustrates the exposure necessary for a handful of targets for spin-independent couplings. A critical feature of the neutrino fog is that it will move to lower cross section if uncertainties in the neutrino fluxes are reduced, opening up new space for continuing searches. 

\section{Landscape of particle dark matter theory}
\label{sec:theory}


The scope of particle dark matter models has evolved since the 1970's when the necessity for Beyond Standard Model (BSM) physics to account for the dark matter in the Universe became evident, resulting in a plethora of diverse ideas today. In the 1980's, BSM models were largely motivated by solving other problems of the Standard Model (SM), containing dark matter candidates almost as an afterthought. This is, for example, the case of supersymmetric versions of the SM (SSM), of which many non-minimal variations (NMSSM) are at the moment compatible with existing experimental limits. In the 1990's, the inclusion of good dark matter candidates became essentially mandatory for all proposed BSM models. At that time, most of the attention focused on the case where the dark matter interactions are feeble due to the exchange of heavy mediators. Since the 2000's and onward, a paradigm shift occurred in which many dark matter models are proposed to fit hints from experiments and/or predict novel dark matter signatures and experiments, with less attention to their completeness, although many nonetheless have implications at accelerators, e.g. searching for light mediators and/or displaced vertices. This has blossomed into exploring many types of dark interactions, and even whole “dark sectors” connected to the SM through “portals” leading to very small couplings with photons, neutrinos, or the Higgs boson. 

\begin{figure}[t]
\begin{center}
\includegraphics[width=1.0\columnwidth]{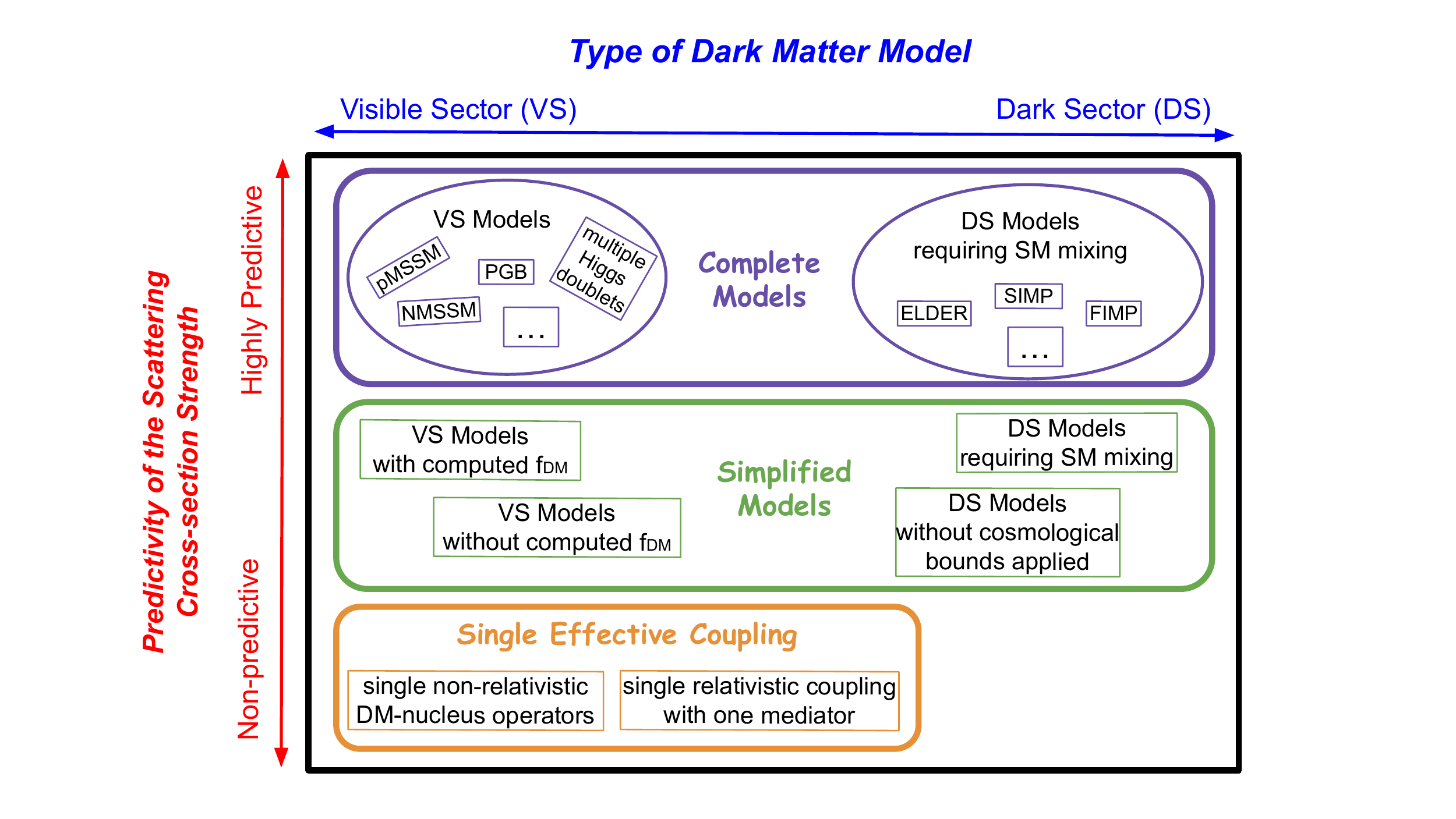}
\caption{Landscape of models classified in terms of predictivity of scattering cross sections strength with nuclei or electrons (vertical axis) and from mostly built within the visible sector or a dark sector (horizontal axis), as explained in the text.  Complete models are highly predictive, single effective coupling models are non-predictive, and simplified models are somewhere in between, in terms of predictivity. Examples of models that predict cross sections at the neutrino fog level are given in the text (FIMP or feebly interacting massive particles models typically predict  cross sections much below the neutrino fog level in the dark matter mass range of interest here).  
}
\label{fig:lanscape_theory}
\end{center}
\end{figure}

 Many of these models predict scattering cross sections with nuclei and/or electrons close to the neutrino fog in multi-ton direct detection experiments, although the magnitude of the cross section often spans a wide range. In the following, we classify models in terms of this feature as schematically shown in Fig.~\ref{fig:lanscape_theory} and mention a few of each type which predict scattering cross sections close to the neutrino fog. We restrict our discussion to only examples in which regions of the scattering cross section versus mass are presented by the authors, some of which we reproduce in our figures. Overviews of these models can be found e.g. in \cite{Battaglieri:2017aum}  or \cite{Billard:2021uyg}. 
 
 The relevance of the neutrino background depends on the  dark matter scattering spectrum. The neutrino fog for dark matter scattering off nuclei we mention in this section refers to Spin Independent (SI) interaction and we reproduce in the figures its level for xenon~\cite{Billard:2013qya,Ruppin:2014bra}.  The effect of the neutrino background is much less pronounced for other types of interactions with momentum suppressed elastic  scattering cross sections and for inelastic scattering (see e.g.~\cite{Gelmini:2018ogy}). The spectrum of dark matter scattering off electrons is always different from the neutrino background spectrum. However this background still has an impact in the discovery reach that depends on the exposure~\cite{Essig:2018tss,Wyenberg:2018eyv}.  

\begin{figure}[t]
\begin{center}
\includegraphics[width=0.80\columnwidth]{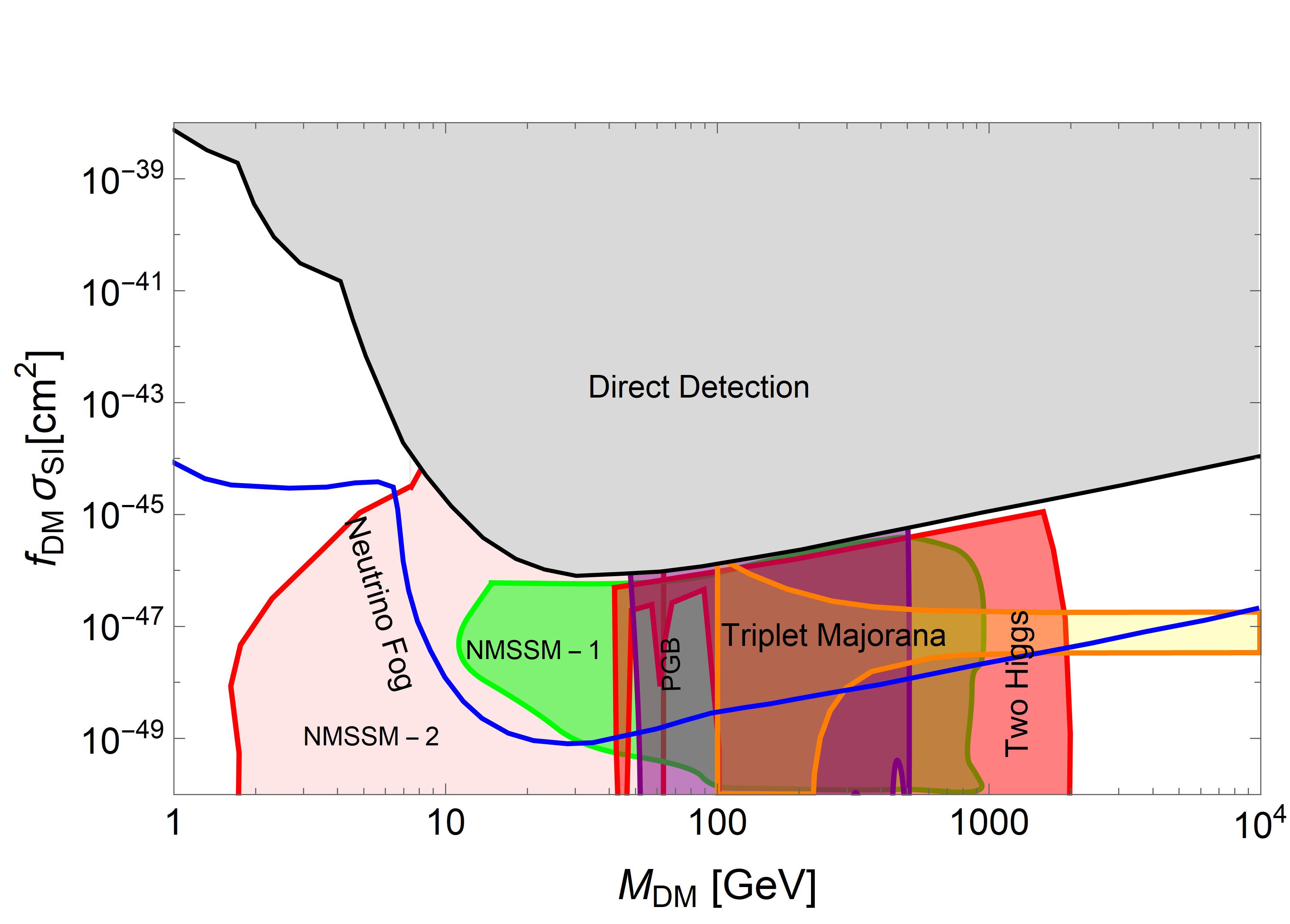}
\caption{Plots of dark matter-proton cross section $\sigma_{\rm SI}$  times dark matter fraction $f_{\rm DM}$ versus particle mass $M_{\rm DM}$ for SI scattering cross sections off nuclei. Examples of predicted regions in several visible sector models are reproduced: the NMSSM models of~\cite{Lopez-Fogliani:2021qpq} (``NMSSM-1", in green) and of \cite{Wang:2020xta} (``NMSSM-2", in pink), the two Higgs doublet model of~\cite{Cabrera:2019gaq} (``Two Higgs", in red), the pseudo-Goldstone boson  of~\cite{Alanne:2020jwx} (``PGB", in purple) and the electroweak triplet Majorana dark matter model of~\cite{Chen:2018uqz} (``Triplet Majorana", in yellow - however note that the relic abundance for this model was not computed). The gray region is excluded by existing direct detection limits~\cite{Evans:2017kti} and the blue line indicates the neutrino fog level in xenon~\cite{Billard:2013qya,Ruppin:2014bra}. Regions predicted by other examples of models mentioned in the text~\cite{VanBeekveld:2021tgn, Mukherjee:2022kff, Khater:2021wcx, Chen:2018uqz, Chen:2019gtm,Berlin:2015ymu} overlap with those shown and were not included for clarity.}
\label{fig:sample_SI-VisibleSector}
\end{center}
\end{figure}
 
 We classify dark matter models as either ``visible sector" or ``dark sector".
 Visible sector models include mostly particles that carry quantum numbers of the SM gauge group. They consist of extensions of the SM,  such as those supersymmetric or with multiple Higgs fields. By contrast, dark sector (or hidden sector) models contain many particles that do not interact with the SM, or interact only through a small mixing with it, often referred to as a ``portal".  The particles of the dark sector typically interact among themselves through new forces (Abelian or non-Abelian gauge groups) which could be spontaneously broken or confining.
 A dark sector could in principle be limited to gravitational interactions with the SM particles, but in this case there is no clear way to ensure that its relic density will appropriately match cosmological observations.  For this reason, the
 portal interactions of a dark sector model are typically essential to determine the dark matter abundance. Given the rich spectrum of the possible production mechanisms, the range of favored dark matter masses is typically wider in dark sector models than in visible sector ones. 
 
 In complete visible sector models, all scattering cross sections with the SM particles can be computed exactly, and cosmological and experimental limits applied to them. Some examples of this type are NMSSM models of~\cite{Lopez-Fogliani:2021qpq} and~\cite{Wang:2020xta}, the Phenomenological Minimal Supersymmetric Models (pMSSM) of~\cite{VanBeekveld:2021tgn} and~\cite{Mukherjee:2022kff}, the multiple scalar doublet models of~\cite{Cabrera:2019gaq} and ~\cite{Khater:2021wcx}, the $U(1)_{L\mu -L\tau}$ gauge SM extension model of~\cite{Singirala:2021gok}, the stable neutral heavy Dirac fermion dark matter (RHN model) in~\cite{Barger:2008qd}, 
 and the pseudo-Goldstone Boson dark matter model of \cite{Alanne:2020jwx}, in which the dark matter consists WIMPs of mass in the GeV to 1.5 TeV range with Spin Independent (SI) or Spin Dependent (SD)  scattering  cross sections off of nuclei.  
 In some visible sector models detailed predictions of the cross section of scattering of the dark matter off nuclei were made while remaining agnostic about how the dark matter relic density could be produced in the early Universe (e.g. for Higgsino like or Wino like fermions of 0.1 to 10 TeV mass in \cite{Chen:2018uqz} and \cite{Chen:2019gtm}). This approach can be motivated by our lack of knowledge of the cosmology during the epoch in which the DM relic abundance in these models is produced (see e.g.~\cite{Gelmini:2006pw, Gelmini:2006pq, Gelmini:2006mr, Berger:2020maa, Dienes:2021woi, Howard:2021ohe}).  In this case, one should be aware that different assumptions about the dark matter production would typically select only a small portion of the parameter space considered.
 
\begin{figure}[h]
\begin{center}
\includegraphics[width=0.80\columnwidth]{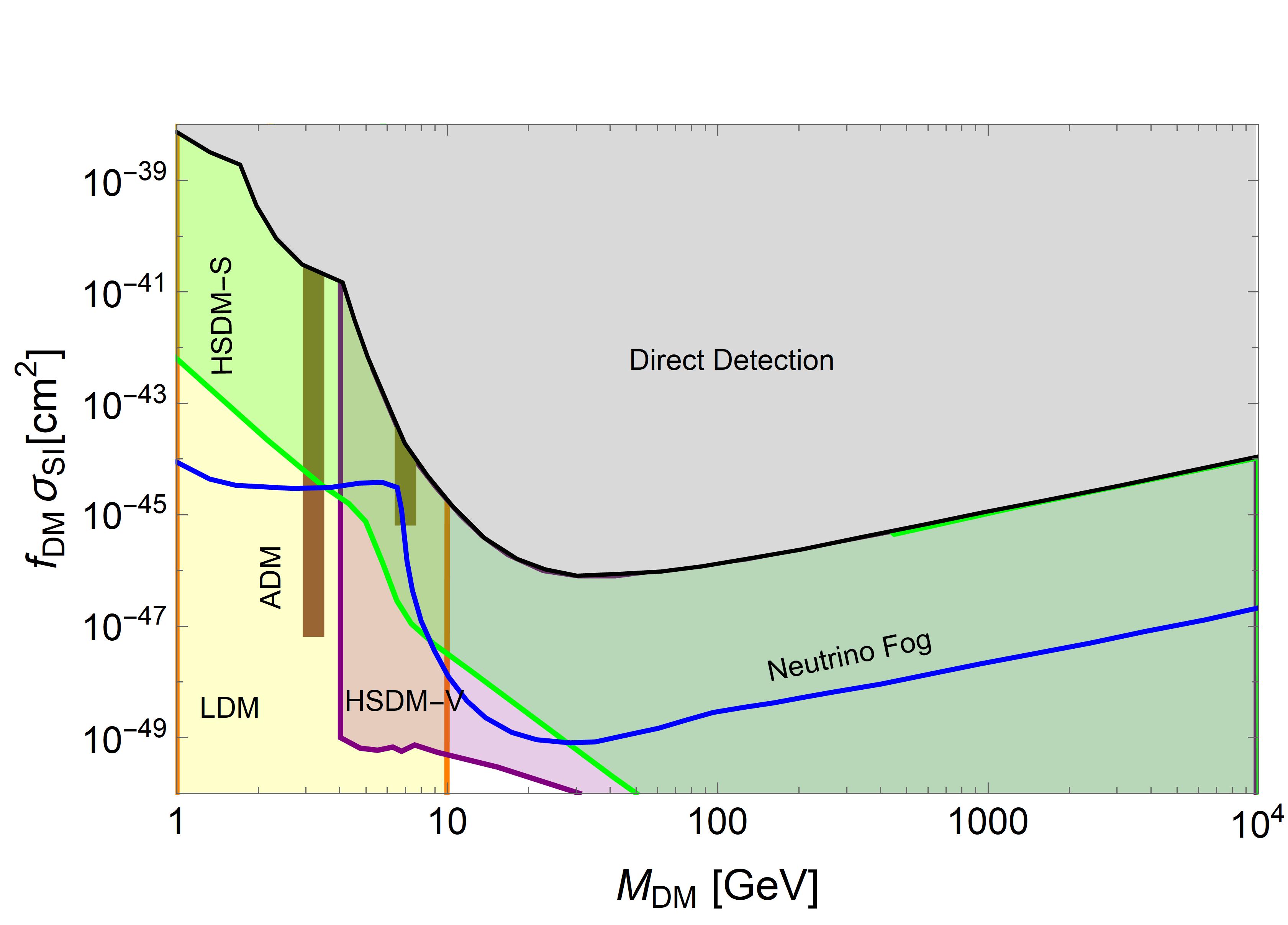}
\caption{Same as in Fig.~\ref{fig:sample_SI-VisibleSector} for SI cross sections but for a few dark sector models: the minimal hidden sector dark matter models of~\cite{Evans:2017kti} with a vector portal (``HSDM-V", in purple) and with a Higgs scalar portal (``HSDM-S", in green), the asymmetric dark matter model of~\cite{Cohen:2010kn} (``ADM", brown vertical bars) and the symmetric and asymmetric light dark matter models of~\cite{Lin:2011gj} (``LDM", in yellow).}
\label{fig:sample_SI-DarkSector}
\end{center}
\end{figure}

 In Fig.~\ref{fig:sample_SI-VisibleSector} we reproduce some of the predictions for SI scattering cross sections (in plots of dark matter-proton cross section versus dark matter mass)  for the visible sector models of Refs.~\cite{Lopez-Fogliani:2021qpq, Wang:2020xta, Cabrera:2019gaq, Alanne:2020jwx, Chen:2018uqz}. Examples of predictions of dark sector model models for SI scattering  are shown in Fig.~\ref{fig:sample_SI-DarkSector}, as explained below. 

\begin{figure}[t]
\begin{center}
\includegraphics[width=0.80\columnwidth]{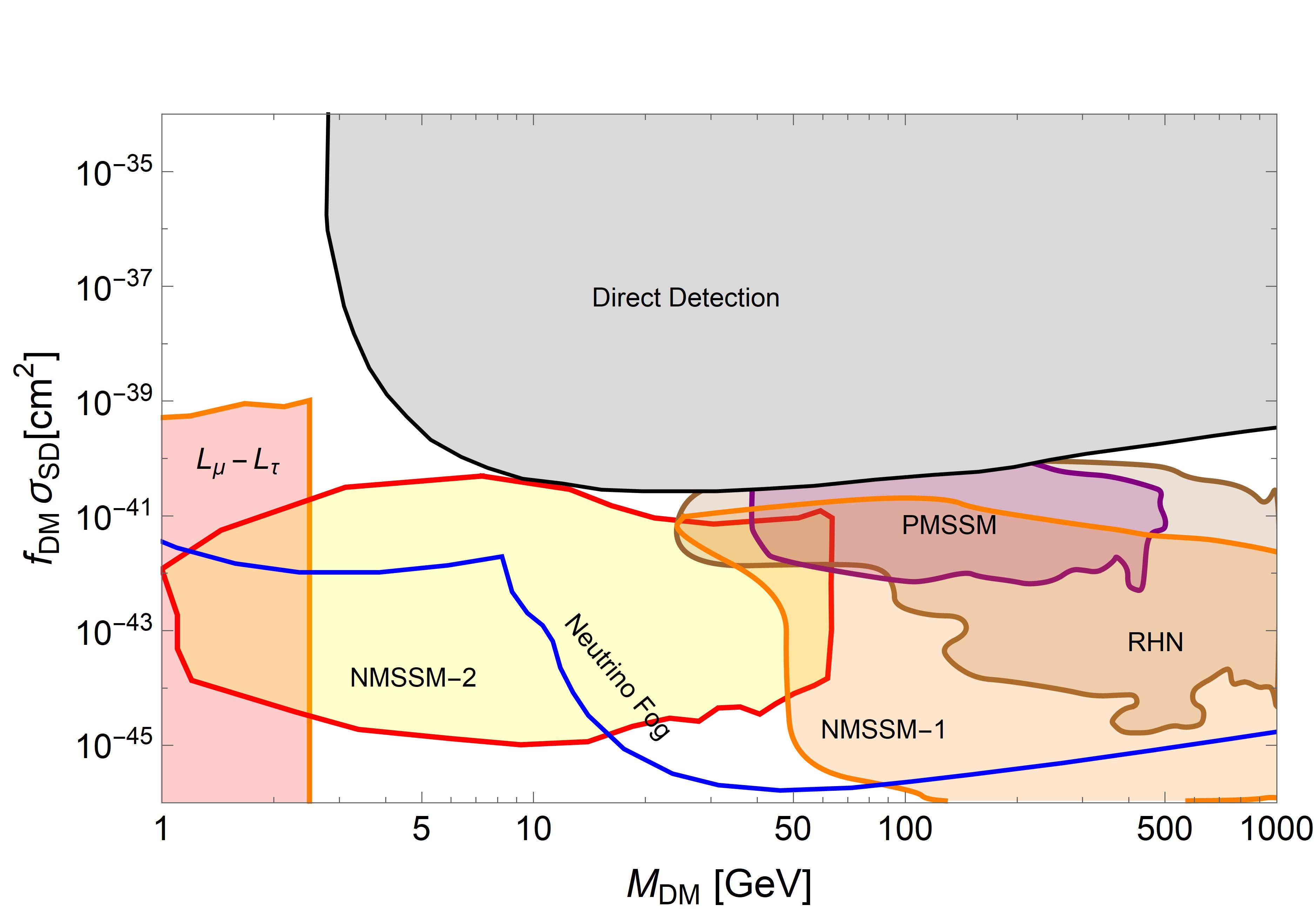}
\caption{Plots of dark matter-proton cross section $\sigma_{\rm SD}$  times dark matter fraction $f_{\rm DM}$ versus particle mass $M_{\rm DM}$ for SD scattering cross sections off nuclei. Examples of predicted regions in several visible sector models are reproduced: the NMSSM models of~\cite{Lopez-Fogliani:2021qpq} (``NMSSM-1", in orange) and of \cite{Wang:2020xta} (``NMSSM-2", in yellow), the pMSSM model of~\cite{VanBeekveld:2021tgn} (``pMSSM", in purple), the $U(1)_{L\mu -L\tau}$ gauge SM extension model of~\cite{Singirala:2021gok} (``$L_\mu - L_\tau$", in pink), the heavy Dirac fermion model in~\cite{Barger:2008qd} (``RHN", in brown). The gray region is excluded by existing direct detection limits and the blue line indicates the neutrino fog level in for C$_3$F$_8$~\cite{Ruppin:2014bra}.}
\label{fig:sample_SD-VisibleSector}
\end{center}
\end{figure}

In Fig.~\ref{fig:sample_SD-VisibleSector} we reproduce some of the predictions for SD scattering cross sections (in plots of dark matter-proton cross section versus dark matter mass)  for the visible sector models of Refs.~\cite{Lopez-Fogliani:2021qpq, Wang:2020xta, VanBeekveld:2021tgn, Singirala:2021gok, Barger:2008qd}.

\begin{figure}[t]
\begin{center}
\includegraphics[width=0.80\columnwidth]{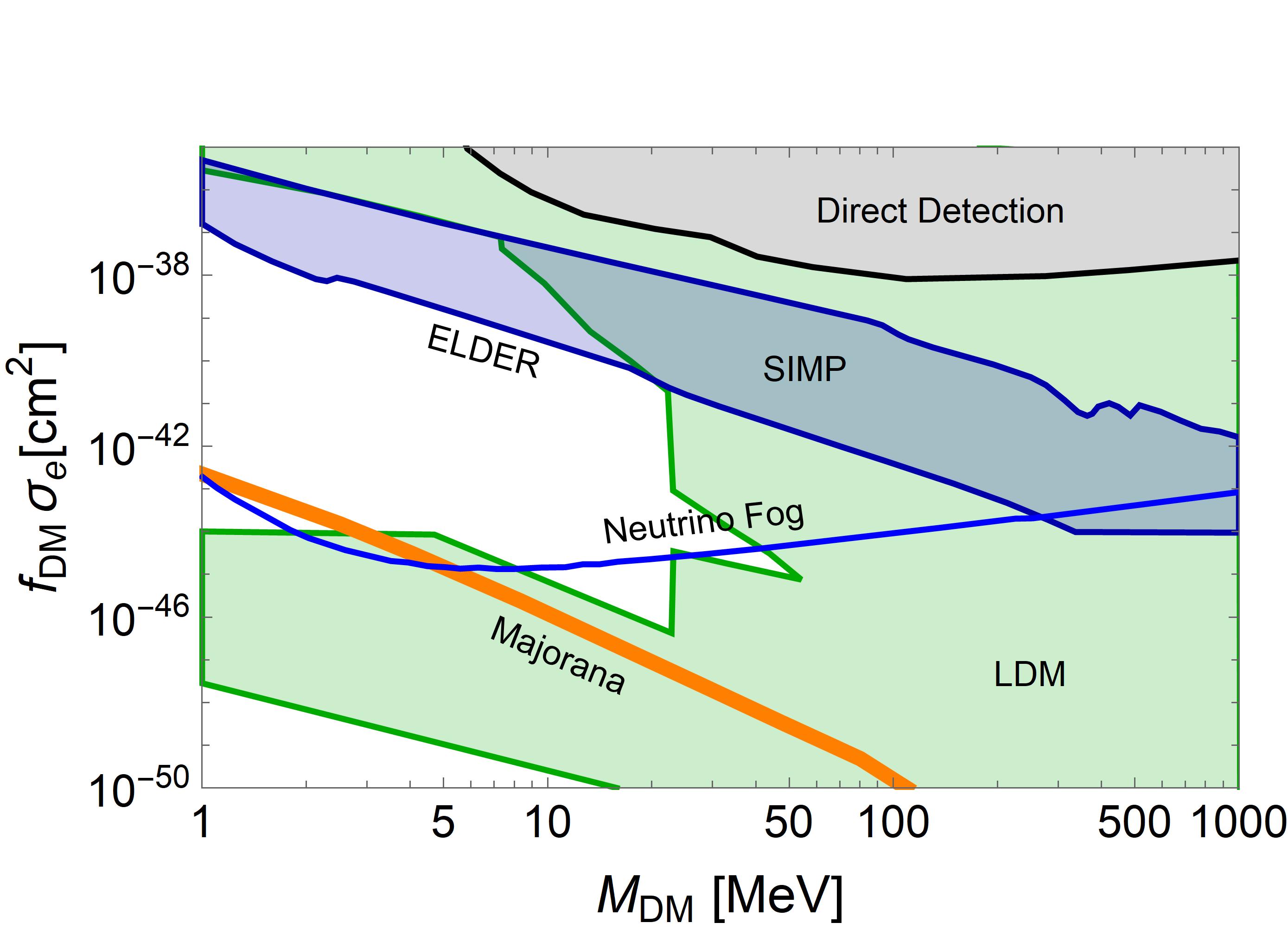}
\caption{Dark matter-electron scattering cross section $\sigma_e$ times dark matter fraction $f_{\rm DM}$ versus particle mass $M_{\rm DM}$. Examples of predicted regions in several models are reproduced: the
$f_{\rm DM}=1$ line for the ELDER model and region for the SIMP model of~\cite{Kuflik:2017iqs} (in blue), the Majorana dark matter line of Fig 6 of~\cite{Battaglieri:2017aum} (in orange), the symmetric and asymmetric light dark matter (LDM) region of~\cite{Lin:2011gj} (in green). The 100\% discrimination $F_{\rm DM } = 1$ neutrino fog level for a $10^5$ kg-y exposure of~\cite{Wyenberg:2018eyv}(blue line) and the region rejected by present direct detection constraints (in gray) are shown for comparison.}
\label{fig:sample_electron}
\end{center}
\end{figure}

 Complete dark sector models exist in which the dark matter interaction with the visible sector is required as an integral ingredient to successfully predict the dark matter abundance. One of the most predictive is the ``Elastically Decoupling Relic" (ELDER) dark matter model \cite{Kuflik:2015isi, Kuflik:2017iqs}, a thermal relic whose present density is determined primarily by the cross-section of its elastic scattering off SM particles. Assuming this scattering is mediated by a kinetically mixed dark photon, the ELDER model makes concrete predictions for scattering off electrons. In this model the dark matter has strong number-changing self-interactions.  This is also the case for the “Strongly-Interacting Massive Particle” (SIMP) model of~\cite{Hochberg:2014dra}, where the relic abundance is determined by the strong self interactions, with the coupling to the visible sector necessary to maintain thermal equilibrium between the SM and dark sector. Without thermal equilibrium, 3 to 2 annihilation would heat up the DM through ``cannibalization"~\cite{Carlson:1992fn,deLaix:1995vi,Boddy:2014yra}, resulting in large velocities incompatible with observations of structure formation in the Universe. ELDER and SIMP model predictions are shown in Fig.~\ref{fig:sample_electron}.
 
Simplified models are defined by a small number of new particles and their interactions, often restricted to a single interaction channel with the SM that is assumed to dominate, while remaining agnostic about others that would be likely to appear in a more complete model.  Some simplified models take into account all cosmological, astrophysical and experimental bounds that apply to them, e.g. those mentioned in this paragraph. 
In the asymmetric dark matter (ADM) from a GeV hidden sector model of~\cite{Cohen:2010kn}, the scattering off nuclei is computed for a few illustrative dark matter  masses -- for a mass of 3.3 GeV, the scattering cross section is close to the neutrino fog (as shown in Fig.~\ref{fig:sample_SI-DarkSector}). 
 In the symmetric and asymmetric dark matter dark sector models of~\cite{Lin:2011gj}, a dark matter particle of mass in the 1 MeV to 10 GeV range could scatter off nuclei (covering all the allowed region shown in Fig.~\ref{fig:sample_SI-DarkSector}) or electrons (as shown in Fig.~\ref{fig:sample_electron})  with cross sections close to the respective neutrino fog levels. 
 In the minimal hidden sector dark models of~\cite{Evans:2017kti}, dark matter with SI scattering off nuclei at the neutrino fog level can have masses in the 0.5 GeV to 10 TeV range (see Fig.~\ref{fig:sample_SI-DarkSector})
In the visible sector Majorana fermion dark matter models of~\cite{Berlin:2015ymu}, SI scattering off of nuclei is predicted at loop level to lie at or just below the neutrino fog for dark matter masses from 0.1 to 1 TeV. 
In~\cite{Battaglieri:2017aum}  a specific example of a Majorana fermion scattering off electrons is given whose cross section  would encounter the neutrino fog (the corresponding line in Fig.~6 of~\cite{Battaglieri:2017aum} is reproduced in Fig.~\ref{fig:sample_electron}).

  Fig.~\ref{fig:sample_electron}  shows the ELDER line and SIMP region predicted in~\cite{Kuflik:2017iqs} for scattering off electrons of 1 MeV to 1 GeV dark matter particles and the region of models of~\cite{Lin:2011gj}, the symmetric and asymmetric light dark matter (LDM) region of~\cite{Lin:2011gj} and the Majorana dark matter line of Fig 6 of~\cite{Battaglieri:2017aum} (in orange),   The 100\% discrimination $F_{\rm DM } = 1$ neutrino floor level for a $10^5$ kg-y exposure of~\cite{Wyenberg:2018eyv} is shown in the figure for comparison (higher fog levels for scattering off electrons corresponding to smaller exposures where computed in~\cite{Essig:2018tss}).

 Dark matter particles could also interact with SM particles through electromagnetic multipole moments, leading to scattering with a very different momentum transfer dependence from the usual SI or SD cases.   Many models realize this idea for fermionic or vectorial dark matter. E.g. the particular vectorial DM model of~\cite{Hisano:2020qkq} shows predictions for scattering cross sections which extend to and into the neutrino fog.  However, one should bear in mind that the relic dark matter abundance and cosmological bounds were not computed.  
 
Finally, we mention two approaches to modeling DM interactions which are completely agnostic with regard to the size of the scattering cross section. 
In the first, a single interaction channel between the DM and the SM mediated by a single messenger is considered (see e.g.~\cite{Gelmini:2018ogy}).  Typically such simplified models consist of Lagrangian terms defining the messenger couplings to the DM and to quarks and/or electrons and a type and mass of the single mediator. The simplest incarnations thus are defined by four parameters: the two coupling constants, the DM mass, and the mediator mass,
with the advantage that the spectrum and experimental signals are relatively easy to understand.  The disadvantage is that in more realistic models, there are often several relevant messengers and/or interactions, whose relative importance often varies across the parameter space and for the various experimental searches (see e.g.~\cite{Profumo:2013hqa}).

A second approach is that of studying single non-relativistic effective field theory (NR EFT) of dark matter-nucleus coupling, a theoretical framework describing different types of nuclear response to dark matter scattering which yields insight into different viable dark matter couplings~\cite{Fitzpatrick:2012ix}. However, most of the dark matter-nuclei interactions defined in terms of a field theoretical Lagrangian formalism involve complex linear combinations of EFT operators in the non-relativistic limit, with the relative importance of each EFT operator weighted by nuclide-specific factors, and realistic relativistic constructions typically require several NR EFT operators for an accurate description (see e.g.~\cite{Gresham:2014vja}).

The examples we provide show that a large variety of dark matter models predict scattering cross sections at or below the neutrino fog.

\section{Experimental sensitivity}
\label{sec:experiment}

\subsection{Prospects for reaching the neutrino fog}

\subsubsection{Currently-funded experiments and R\&D toward next generation}
\label{sec:currentexperiments}

\paratitle{Liquid Xenon} 
%
Liquid Xenon (LXe) is an ideal target for rare event detection, and is particularly well suited for WIMP searches. Xenon is an excellent scintillator, highly transparent to its own scintillation light. It is readily available on the market with a typical world production rate of about 65 tonnes/year. Xenon can be effectively purified for particle detector applications in the multi-tonne range. The high density of LXe allows for the construction of compact detectors that feature a low-background inner core 
exploiting self-shielding
due to the large atomic number of xenon and the absence of significant long-lived radioisotopes. 
When combined with a technology capable of position reconstruction, namely a two-phase Time Projection Chamber (TPC), this allows the definition of a low-background inner fiducial volume that is used for rare event searches.

When conducting searches for low energy nuclear recoil events Xe TPC detectors have demonstrated discrimination powers, rejecting single-site electron recoil backgrounds, of $>$99.7\% down to nuclear recoil equivalent energies 2\,keV~\cite{LUX:2020car}. Lower analysis thresholds can be achieved by dropping the requirement of the presence of a S1 (primary scintillation) when operated in S2-only (ionization) mode.

\begin{figure}[!htbp]
\begin{center}
\includegraphics[width=0.48\columnwidth]{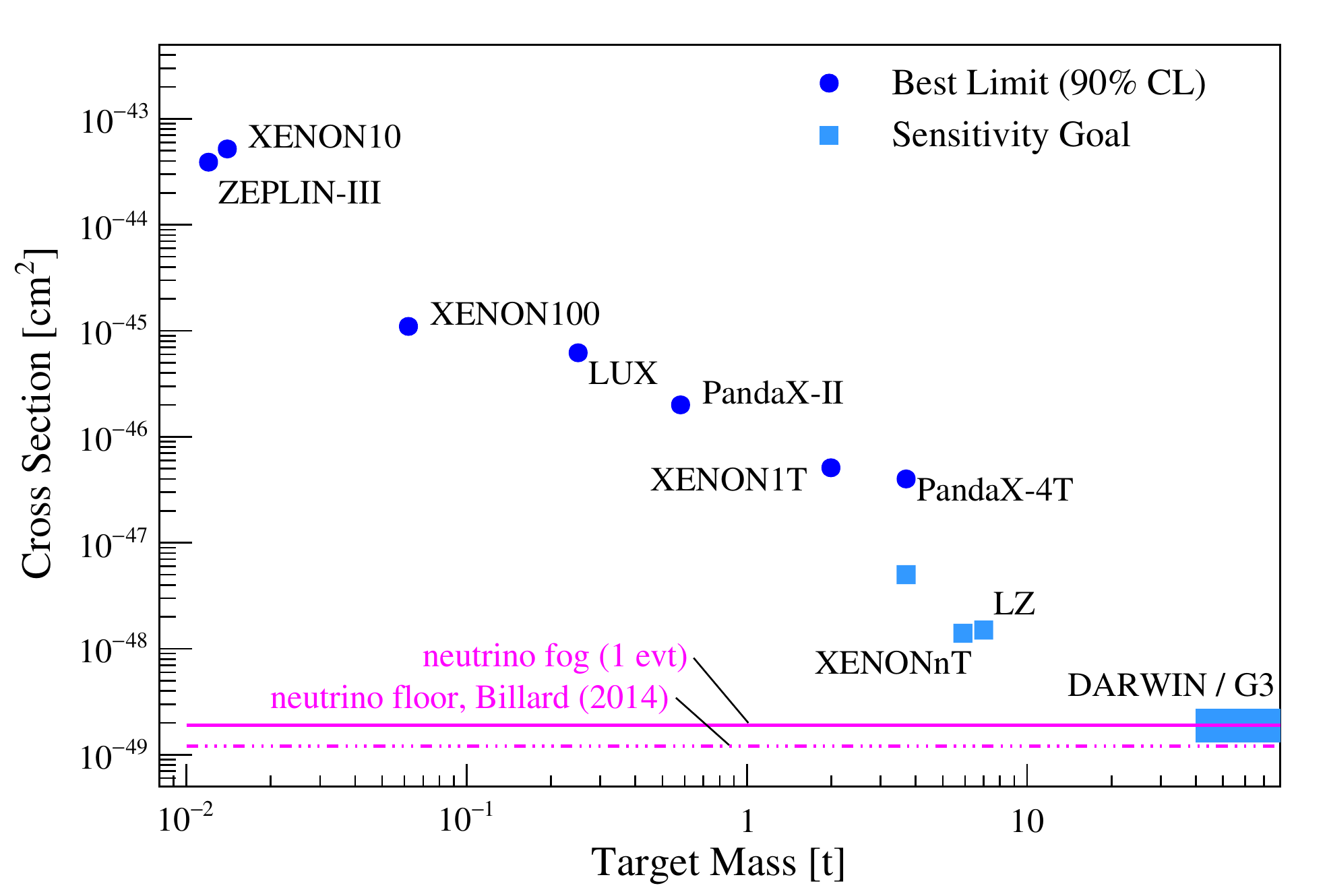}
\includegraphics[width=0.48\columnwidth]{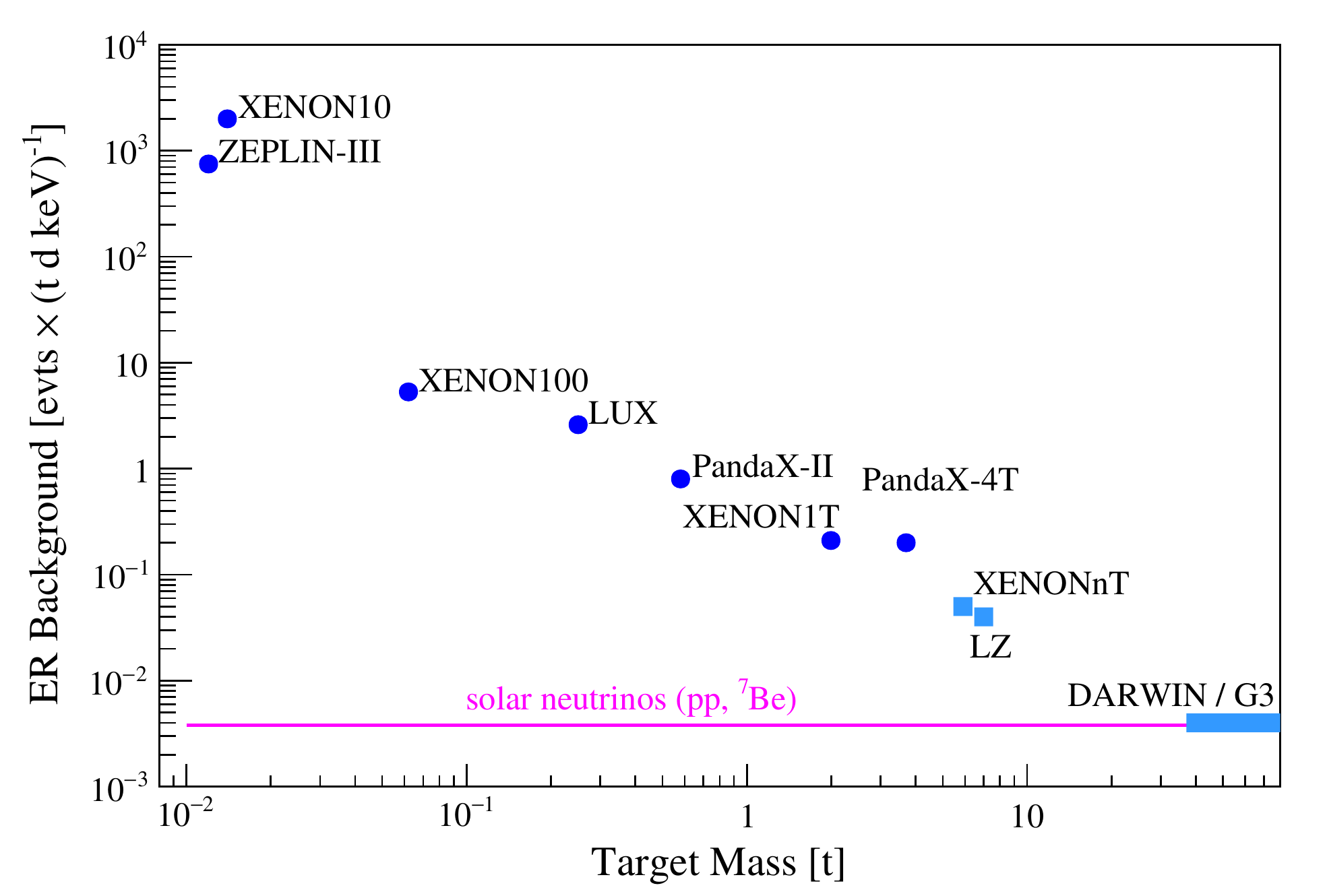}
\caption{(left) Development of LXe-TPC technology. The plot shows the improvement in sensitivity to spin independent WIMP-nucleon coupling (for a mass of 50~Gev/c$^2$) achieved by LXe experiments of increasing target masses. 
Sensitivity goals are also reported for experiments that have not yet been completed. 
(right) The plot shows, as function of the target mass, the progress made in terms of background suppression. 
Cross section values and background rates are extracted from Ref.~\cite{PhysRevLett.100.021303,PhysRevD.94.122001,AKIMOV201214,PhysRevLett.116.161301,Wang_2020,XENON:2018voc,PandaX:2018wtu,PandaX-4T:2021bab,AKERIB201804,XENON:2020kmp}.
}
\label{fig:xenon_evolution}
\end{center}
\end{figure}

Figure~\ref{fig:xenon_evolution} summarizes the impressive achievements of the LXe-TPC technology, pioneered in early 2000s by XENON-10~\cite{XENON10:2007prx}, ZEPLIN-II~\cite{ALNER2007287}, and ZEPLIN-III~\cite{Akimov:2011tj}. Since then a series of LXe-TPCs of increasing target mass and decreasing background has led the direct detection field across more than 3 orders of magnitude, passing through XENON100~\cite{APRILE2012573}, LUX~\cite{AKERIB2013111}, PandaX-I, and PandaX-II~\cite{pandax} with  XENON1T~\cite{xenon1t,XENON:2018voc} and PandaX-4T~\cite{PandaX-4T:2021bab} presently holding the most stringent constraints for WIMP masses above 2\,GeV/c$^2$ (0.1\,GeV/c$^2$ if the Migdal effect is assumed) in a 1\,tonne$\cdot$year and 0.6\,tonne$\cdot$year exposure, respectively~\cite{PhysRevLett.123.251801, PhysRevLett.123.241803,PandaX-4T:2021bab}.


Two new multi-tonne LXe-TPCs, LZ~\cite{AKERIB201804} and XENONnT~\cite{XENON:2020kmp}, have been operating since 2021 with a target sensitivity of about 1-2$\times$10$^{-48}$ cm$^2$ (see Fig.~\ref{fig:xenon_evolution}).
The US-European-Japanese collaboration XENON is running the XENONnT experiment at the Laboratori Nazionali del Gran Sasso (Italy). The detector features a 5.9\,tonne LXe target (4.0\,tonne expected fiducial) surrounded by a Gd-doped water Cherenkov active veto to suppress and tag neutron-induced background.
The experiment plans to accumulate a 20\,tonne$\cdot$year exposure. The projected neutrino-induced background rate in the WIMP ROI is of 0.27 events/tonne$\cdot$year and 0.08 events/tonne$\cdot$year for solar and atmospheric neutrino respectively, with the former not affecting WIMP searches for masses larger than 10\,GeV/c$^2$. 

The US-European LUX-ZEPLIN (LZ) experiment is presently operating at Sanford Underground Research Facility (USA) and aims to accumulate a 15\,tonne$\cdot$year exposure~\cite{AKERIB201804}. The detector features a 7\,tonne target (5.6\,tonne expected fiducial), surrounded by an instrumented LXe skin layer and a Gd-loaded liquid scintillator which together form an efficient anti-coincidence veto and in~situ background monitor. 
The expected neutrino-induced background rate in the WIMP signal region is 2.35\,events/tonne$\cdot$yr (solar) and 0.02\,events/tonne$\cdot$yr (atmospheric).

In China, PandaX-4T~\cite{Kang:2010zza,PandaX:2018wtu,PandaX-4T:2021bab}
is expected to continue operations until 2025 and then to be incrementally upgraded into a multi-ten-ton experiment with a nominal fiducial target of 30~ton.

The DARWIN collaboration~\cite{DARWIN:2016hyl} has already started in Europe an intense R\&D program for a 40-80 tonne Xe target. Given their current leadership roles and expertise, the US teams in both XENONnT and LZ (about 50\% of the community) are well-positioned to contribute significantly to this next-generation effort.   
In 2021, scientists in the LZ and XENON/DARWIN collaborations formally expressed their intent to join forces towards towards this next-generation xenon-based experiment, pursued by a single, joint scientific collaboration~\cite{mou}.
There would be substantial opportunity cost in delaying the US engagement.

Two decades of intensive development and deployment has made the LXe-TPC an extremely mature technology capable of being extended to a larger scale with reliable sensitivity projections and consistently solid progress (see Fig.~\ref{fig:xenon_evolution}). 
An LXe-based experiment targeting fiducial exposures on the order of 1\,ktonne$\cdot$year would fully explore the accessible WIMP parameter space as constrained by the neutrino fog. This assumes particle masses $\gtrsim 5~$GeV/c$^2$ and current detector energy thresholds. The sensitivity for an exposure of 1\,ktonne$\cdot$year delivering a $3\sigma$ dark matter particle discovery~\cite{Aalbers:2022dzr}, likely achieved with a LXe instrument with a fiducial mass above 100~tonnes, is illustrated in Fig. \ref{fig:limitplot_SI}. 
The strategy to approach such an exposure will be informed by outcomes from the current projects, complemented by a well-defined set of R\&D activities that should take place concurrently. 

The community support for a next generation xenon-based experiment is very broad. A 600-author white paper outlining the physics reach of such an experiment was recently published~\cite{Aalbers:2022dzr}. The large detector scale and low background, combined with many advantageous properties of xenon properties as target, enable a rich science portfolio that extends beyond WIMPs, transforming such a detector into a cost-effective, broad, low-background astroparticle physics observatory.

\paratitle{Liquid Argon} 
The successful campaign of the DEAP-3600 experiment demonstrates the feasibility of multi-tonne-scale LAr detectors, including the development and use of ultra-high purity acrylic~\cite{amaudruzDesignConstructionDEAP36002019a} and novel techniques for achieving heavily suppressed Rn contamination. Thanks to its large size, DEAP-3600 also demonstrated the promise of large-scale LAr detector for searching for dark matter at the highest masses, up to the Planck scale~\cite{deapcollaborationFirstDirectDetection2022}.  DarkSide-50 demonstrated the sensitivity of a LAr TPC using Underground Argon (UAr), heavily depleted in $^{39}$Ar, to light dark matter with nuclear~\cite{darksidecollaborationLowMassDarkMatter2018} and electronic~\cite{thedarksidecollaborationConstraintsSubGeVDarkMatter2018} couplings. This is due in part to the relatively light argon nucleus and to the high degree of chemical purity achievable in LAr due to its cold temperature.  The DarkSide-LowMass experiment is now being planned to leverage these advantages to search for dark matter with cross sections down to the solar neutrino fog with an estimated tonne-year exposure of further-depleted UAr.

The DarkSide-20k (DS-20k) experiment of the multinational Global Argon Dark Matter Collaboration (GADMC) will search for dark matter using an UAr target instrumented as a dual-phase time projection chamber (TPC)~\cite{Aalseth:2018gq}.  The GADMC includes more than 400 scientists from over 60 institutions, mostly coming from the ArDM, DarkSide-50, DEAP-3600, MiniCLEAN, and XENON collaborations, and DS-20k will inherit successful elements from these experiments. These include the use of UAr, which DarkSide-50 demonstrated to have an $^{39}$Ar concentration at least 1400 times smaller than atmospheric argon~\cite{Agnes:2015gu,Agnes:2016fz,Agnes:2018ep}, and a radiopure acrylic structure, a technology pioneered by the DEAP-3600 experiment~\cite{Nantais:2013jp,Amaudruz:2018gr}.  The TPC is designed to take advantage of the favorable properties of liquid argon, including demonstrated electron recoil background discrimination power better than $10^8$~\cite{Adhikari:2021} and excellent chemical purity~\cite{Ajaj:2019jx,Ajaj:2019jk}, and operate with $<0.1$ background events within the 20.2~t fiducial volume over a ten year run, other than an expected ($3.2 \pm 0.6$) events from coherent neutrino scattering.  The DS-20k experimental apparatus consists of three nested detectors installed within a membrane cryostat nearly identical to the two existing ProtoDUNE cryostats~\cite{Abi:2020gv,Abi:2020hj,Abi:2020je}: the inner two-phase argon TPC, a neutron veto, and an outer muon veto. The apparatus will be located in Hall C of the Gran Sasso National Laboratory (LNGS). The DS-20k UAr target will be extracted by Urania~\cite{Aalseth:2018gq}, an argon extraction plant capable of extracting 330~kg/d of UAr, and purified with the Aria plant~\cite{Agnes:2021us}, a 350~m cryogenic distillation column designed to separate argon and other rare stable isotopes.  As illustrated in Fig.~\ref{fig:limitplot_SI}, DarkSide-20k experiment extends the cross section vs. mass range sensitivity in the search for dark matter to  $4.6\times 10^{-48}$\,cm$^2$ for a 90\% C.L. exclusion (and $1.5\times 10^{-47}$\,cm$^2$ at a $5\sigma$ discovery significance) for a 1\,TeV/$c^2$ WIMP, well beyond any current or presently funded experiment.  This will lead to either discovery, confirmation, or exclusion of the WIMP dark matter hypothesis down to the level where coherent scatters from atmospheric neutrinos become an irreducible background.  DarkSide-20k will also be sensitive to a galactic supernova neutrino burst originating anywhere in the Milky Way Galaxy~\cite{Agnes:2021ke}.

The ultimate objective of the GADMC is the construction of the Argo detector, which will have a 300~t fiducial mass and will push experimental sensitivity into the atmospheric neutrino fog.  The excellent electron recoil (ER) rejection possible in argon will eliminate backgrounds from solar neutrinos, which will extend the sensitivity of Argo beyond that of technologies with more limited ER discrimination.  Such a large detector would also have excellent sensitivity to a neutrino burst associated with a galactic supernova~\cite{Agnes:2021ke}.  If located at SNOLAB or at similar depth, Argo will also have the potential to observe CNO neutrinos for the first time and solve the Solar Metallicity Problem~\cite{Agnes:2020wd,Franco:2016ex}.  The further expansion of several technologies are critical for the success of Argo. The continued development and operation of Urania~\cite{Aalseth:2018gq} and Aria~\cite{Agnes:2021us} would enable 400~t of UAr to be extracted and purified over a period of about five years. Facilities for the long-term underground storage and assay of this argon are also needed. The Argo detector will be instrumented with more than 100~m$^2$ of photodetectors. This would be greatly simplified by the continued evolution of the large-area SiPM-based photosensors developed for DS-20k~\cite{DIncecco:2018hy,DIncecco:2018fx} into digital SiPMs, which would reduce the quantity of cables, significantly reduce noise, and decrease the background rate in the detector. Finally, all future detectors entering into the neutrino fog would benefit from improved atmospheric neutrino background modeling, which currently dominates the uncertainty on the experimental sensitivity.

\paratitle{Solid State} 
The Super Cryogenic Dark Matter Search (SuperCDMS) SNOLAB experiment is a 2$^{nd}$ generation dark matter experiment, which will commission its solid-state detectors 2 km underground at the SNOLAB facility in Sudbury, Canada.  The solid-state detectors are germanium or silicon crystals patterned with both Quasiparticle-assisted Electrothermal-feedback Transition edge sensors (QETs) as phonon sensors and electrodes used for charge readout and/or biasing the crystal.  The patterned QETs and electrodes are optimized to operate as either High Voltage (HV) or interleaved Z-dependent Ionization and Phonon (iZIP) detector.

In an HV detector, the electrodes are used to apply a relatively large bias voltage across the crystal.  The applied voltage gives rise to an \textbf{E}-field that causes electrons and holes to drift to opposing faces of the detector. As they move through the crystal, the charges scatter off the lattice, generating additional phonons via the Neganov-Trofimov-Luke (NTL) effect ~\cite{Neganov1985, Luke1988}.  The resulting total energy observed by the QETs on the detector faces is $E_{tot} = E_r + N_{eh}eV_{b}$, Where E$_r$ is the initial recoil energy, and N$_{eh}$ is the number of electron-hole pairs initially created. These devices have an improved ultra-high resolution and reach lower thresholds allowing them to probe lower DM masses.  The HV detectors are expected to explore WIMP DM masses down to ~0.3 GeV. 

In an iZIP detector, a low voltage bias is applied in order to minimize the NTL effect. The electrodes are also used as charge collectors.   iZIP detectors therefore measure both prompt phonons and ionization charge signals, which can be used to perform ER/NR discrimination based on the difference in ionization yield. Furthermore, optimization of the electrode layout allows identification of surface events via a charge signal collection asymmetry: bulk (nominally symmetric signals) and surface (highly asymmetric signals) events.  This allows rejection of beta particles and further reduces the background in the operation of these devices. The advanced rejection capabilities of these devices project sensitivities in a ``background-free'' mode to WIMPs with masses $>$5 GeV and in a “limited-discrimination” mode to WIMPS $>$1 GeV~\cite{SCDMS2017}.

The SuperCDMS SNOLAB experiment anticipates observing approximately 50 events associated with neutrino interactions, but will not reach the neutrino fog.  The background and detector improvements needed to reach the fog have been identified as part of a near-term SuperCDMS upgrade plan.  Backgrounds improvements include sourcing new material, replacing components with lower background alternatives, and improving detector fabrication/tower assembly to reduce the $^{210}$Pb plated onto the surface from radon, all within a reasonable cost for implementation.  Proposed detector upgrades common to both HV and iZIP style detectors include: 1) smaller detector sizes; 2) lowering the TES critical temperature (T$_C$); and 3) improving phonon transmission across interfaces.  Scaling to smaller detectors will improve the phonon/ionization resolution of the SuperCDMS detectors, which in turn will improve rejection of bulk ER backgrounds. This development is reasonably mature with prototype Si HVeV detectors (an HV style device) already deployed at test facilities ~\cite{SCDMS2019,SCDMS2020}.  R\&D efforts are set to shift toward the development of a Ge HVeV detector and optimization of these devices.  Lowering of the TES T$_C$ is less mature, but a viable path forward is known. The thin tungsten film that is at the heart of the QET can grow in two different phases, $\alpha$-W with a T$_C$ of 15 mK or $\beta$-W with a T$_C$ over 2 K, during deposition. By mixing these two phases in different ratios the T$_C$ can be tuned between the two extremes.  The challenge is in identifying the proper deposition parameters under which the W film will grow at the proper ratio in a reproducible and controllable manner.  Improvements in phonon transmission from the crystal all the way to the W TES is the least mature advancement being considered, with no immediate avenues forward identified.

\paratitle{Bubble Chambers} 
Bubble chambers present a scalable, background-discriminating technology for dark matter detection with the unique capability to operate with a broad variety of target materials, requiring only that the target be a fluid with a vapor pressure.  Multiple experimental efforts are using this flexibility to push dark matter sensitivity into parameter spaces inaccessible to other detection techniques.

The PICO Collaboration uses freon-filled bubble chambers for nuclear recoil detection in targets with high spin-dependent and low spin-independent cross-sections, allowing the exploration of orders-of-magnitude more dark matter parameter space before reaching the neutrino fog than can be achieved in Si, Ge, and noble-liquid targets.  There is strong physics motivation for freon bubble chambers out to kiloton-year exposures, exposures that are plausible with this technique thanks to its field-leading electron recoil rejection ($O$(1) in $10^{10}$ ER events misidentified as nuclear recoils~\cite{PICO:2019rsv}) and monolithic liquid target.  Past PICO experiments at SNOLAB have reached exposures of $\sim$3~ton-days~\cite{PICO:2019vsc}, including a zero-background (observed) ton-day exposure at 3.3-keV threshold~\cite{PICO:2017tgi}.  The PICO-500 experiment, funded by the Canada Foundation for Innovation, is projected, given conservative estimates for muon-induced neutron rejection, to reach an exposure of  ton-years on a C$_3$F$_8$ target by 2025, including 63 ton-days at 3.2-keV threshold, giving 3 expected solar neutrino CE$\nu$NS events, followed by 126 ton-days above the $^8$B CE$\nu$NS endpoint.  This results in a spin-dependent WIMP-proton sensitivity at the $10^{-42}$~cm$^2$ level, commensurate with the spin-dependent WIMP-neutron sensitivity expected from Generation-2 LXe-TPCs.  Unlike the Generation-2 LXe-TPCs, however, PICO-500's projection is still four-orders-of-magnitude \emph{above} the C$_3$F$_8$ neutrino fog.

A freon bubble chamber large enough to reach atmospheric neutrino sensitivity will require the development of a completely new detector ``inner vessel,'' i.e. the vessel containing the superheated liquid target.  Two factors limit the size of the synthetic silica glass jars used for PICO-500: no facilities exist to construct jars larger than those made for PICO-500, and if larger silica jars were to be constructued, the detector livetime would be limited by the alpha activity of synthetic silica.  The surface chemistry and smooth surface provided by silica glass will need to be replicated with a material capable of surface radioactivity of less than 5 nBq/cm$^2$ in order to construct a 50-ton detector sensitive to an atmospheric neutrino event within 5 years.  This level of radiopurity has been exceeded in the large surfaces of Kamland-ZEN \cite{Gando:2020ggh}.  Tests of similar materials are ongoing to determine their mechanical and surface suitability with published results expected by 2023.  Inner vessel replacement of an existing detector would then follow prior to proposing a kiloton-year scale experiment.

The Scintillating Bubble Chamber (SBC) Collaboration uses liquid-noble bubble chambers to extend the nuclear/electron recoil discrimination capability of freon-filled chambers to the sub-keV thresholds needed to reach the spin-independent neutrino fog at 1~GeV.  Liquid-noble bubble chambers differ from their freon-filled cousins in two (related) ways. First, the scintillation coincident with bubble nucleation in the target allows event-by-event energy reconstruction~\cite{Baxter:2017ozv}, which can be used to reject of backgrounds above the few-keV scintillation detection threshold.  Second, noble liquids can be superheated to a far greater degree than molecular fluids, showing sensitivity to sub-keV nuclear recoils while remaining completely insensitive to electron-recoil backgrounds~\cite{Durnford:2021cvb}.  The combination of scalability, low threshold, and background discrimination \emph{at low threshold} gives the noble-liquid bubble chamber unique capability to explore the neutrino fog in the 1--10~GeV WIMP mass range.

The ultimate threshold reach of the liquid-noble bubble chamber is not yet known.  A 10-kg LAr bubble chamber (also capable of operation with LXe) has been built at Fermilab to resolve this question, designed to probe thermodynamic thresholds as low as 40~eV and accomplish nuclear recoil sensitivity calibrations with $O$(10)-eV resolution.  This device is now being commissioned, and low-threshold calibrations will begin in 2023.  Meanwhile, the Canada Foundation for Innovation has funded the construction of a twin 10-kg device for SBC's first dark matter search \cite{Giampa:2021wte}, which has received GW1 approval at SNOLAB.  At a 100-eV nuclear recoil detection threshold (SBC's benchmark until the calibration campaign is complete), this 10-kg LAr chamber will observe 2.5 solar neutrino CE$\nu$NS events per live year.  A follow-up 100-eV threshold LAr experiment at the scale of PICO-500 (1-ton-year exposure) will be sufficient to explore the neutrino fog to $n=2$ at 1-GeV WIMP mass.

\subsubsection{New technologies}

\paratitle{Super-cooled detectors} 
The snowball chamber is a nascent technology which is analogous in operational principles to superheating in bubble chambers and supersaturation in cloud chambers, except that it relies on supercooling~\cite{szydagis2021}.  The first prototype was constructed by Profs. Levy and Szydagis with students at UAlbany SUNY and shown to likely be sensitive to nuclear recoils from neutrons as the radioactive calibration source, with low sensitivity to electron recoils, as in dark matter bubble chambers~\cite{PICO:2019vsc}. The detector relies on lowering the temperature of liquid water below its freezing point in a sufficiently clean and smooth container so that it becomes metastable instead of immediately solidifying.  An incoming particle such as a dark matter WIMP should be able to trigger the phase transition, and potentially encode directionality as well via the intense hydrogen bonding of water.  Advantages of this technique would be the potential for sub-keV energy threshold and the use of water (for scalability, ease of purification, background neutron moderation, and excellent spin-dependent-proton sensitivity).  If the threshold is indeed as low as claimed for decades for supercooled water in atmospheric sciences, then even only a few kg deployed underground for only a few years could lead to world-leading sub-GeV limits for both the standard SI and SD-proton operators.

In order for this detector technology to become viable, numerous challenges will need to be overcome.  The water volume will need to be sufficiently purified in order to achieve a low energy threshold by lowering the temperature sufficiently without nucleation sites present; reduction in background nucleation through these mitigations, combined with faster reheating methods post-event, should lead to the required livetime of $>$50\%; calibrations of backgrounds from all sources will need to be performed, using betas and gamma-rays to fully characterize the electronic recoil discrimination power, as well as alphas to determine how much Radon contamination would be an issue, and these calibrations would need to be performed as a function of temperature (and pressure) with the goal of finding a “sweet spot” temperature.

With most of the materials, equipment, and supplies exist already through earlier seed-funding initiatives, even for building a larger-scale (only grams tested thus far) viable dark matter experiment that is ready for underground deployment, the required resources for construction would be limited, and the funding agencies will need to take a risk on a new idea that is not only an extrapolation from existing concepts: the cost is low so the risk is low, but the return high, based on the advantages discussed earlier.
\paratitle{Low Background DUNE-like module} 
Work is underway to explore the feasibility of a low background kTon-scale liquid argon time projection chamber in the context of the Deep Underground Neutrino Experiment (DUNE).  The DUNE program consists of modules, and though the designs for the first two modules have been selected, the third and fourth (the `modules of opportunity’) remain to be determined.  A recent community effort~\cite{SnowmassLowBgDUNElikeWP} explored the option of making one of these a dedicated low background module, with possible sensitivity to high mass WIMPs~\cite{Church_2020}.  This detector could also confirm a galactic WIMP signal discovered in the generation two detectors using an annual modulation.  

This DUNE-like detector would adapt the standard vertical drift design, with the addition of an optically isolated inner volume where increased light detection allows improved energy resolution at low energies and pulse shape discrimination for background reduction.  The detector would take advantage of the significant self-shielding of the liquid argon to reduce the backgrounds in a 3 kton fiducial volume (compared to 10 ktons in the full module).  Use of low radioactivity underground argon reduces the argon-39 and argon-42 internal backgrounds.

The primary research and development challenges for this detector design are associated with the large scale; in general, the radioactive background requirements are less strict than dedicated dark matter experiments.  Quality control of an assay program will need to be strict to ensure the large amount of material meets requirements.  Radon purification and emanation control in large amounts of liquid argon will need to be demonstrated.  Cleanliness controls for a large detector assembled underground will need to be developed and demonstrated.  Current known underground argon sources are not large enough to supply a detector of this size, and the collaboration is in discussion with commercial gas producers to determine whether a cost-effective supply can be achieved.  The main engineering challenges of this detector are associated with the production and installation of a large amount of SiPMs required for instrumenting the detector to reach the required energy threshold.
\paratitle{Giant gas TPCs in pressurized caverns}
When can a WIMP search use a gas target rather than a liquid?  Gas targets have some microphysical advantages, including: reduced ionization quenching for low-energy nuclear recoils; track-length-based recoil/beta discrimination; low-Fano-factor calorimetry that resolves x-ray backgrounds as lines. A major disadvantge of gas targets is their lack of self-shielding.  However, at the scales needed to reach the neutrino fog, we believe there a route to ultra-high-pressure gas TPCs.  If we can adopt petroleum-industry standard methods for storing large volumes of high pressure gas underground, we can build ultra-high-pressure WIMP targets (up to 500 bar)---big enough to begin self-shielding---with what we suspect to be low-cost instrumentation.  For one example of this new geometry, we describe a 10~m diameter, 80~m tall cylindrical target balloon, which we fill with 500~T of neon at 100~bar and operate as an inward-drift gas TPC; that configuration was optimized for neutrino physics but appears to offer powerful mid-range (2--20 GeV) dark matter sensitivity.  Using solution mining methods in a salt dome, a cavern big enough to host this (including 20~m of gas shielding) could be excavated for a moderate cost.  Many other options (different gases, pressures, sizes, geometries) appear worth exploring.  There are many physics, mechanical engineering, and cost uncertainties to this approach, but we argue that basic R\&D now will help us identify scalable, low-cost detector configurations using previously-impossible targets.

\subsection{New technologies to push below the neutrino fog}
\label{sec:beyondnufog}
\subsubsection{Motivation for Directional Detection}

There are three predicted nuclear recoil signatures of particle dark matter1: 1) an excess in the observed nuclear recoil rate over the predicted background from radioactive impurities and neutrino scattering; 2) an annual modulation in the nuclear recoil rate and energy spectrum due to the motion of the earth around the sun; and 3) a fixed dipole-shaped angular recoil distribution in galactic coordinates, resulting from the motion of the galactic disk with respect to the dark matter halo -- in a terrestrial detector, this last effect is seen as a DM wind, with a direction that oscillates due to the large tilt angle between the earth's spin axis and the luminous plane of the Milky Way (and solar motion).  A direct detection experiment can pursue one or several of these signatures.

Most leading direct detection experiments to date, such as the liquid noble gas detectors described above, have primarily aimed to detect Signature 1.  Signature 2 was used by DAMA/LIBRA~\cite{Bernabei:2019ajy} to claim a detection of dark matter that has not been confirmed by subsequent experiments.  This second signature can be used even in the presence of backgrounds, but requires exquisite detector stability and a large number (1000s) of observed DM events to detect the relatively small (few-percent-level) modulation magnitude, and is subject to interference from temperature effects and cosmic ray-induced background, which also are expected to have an annual modulation.  Signature 3 requires significantly more complex detectors, capable of reconstructing the directions of low-energy nuclear recoils, not strictly required for Signatures 1 and 2.  On the other hand, because the majority of recoils are expected to point to a single hemisphere in galactic coordinates, the effect is large, so that only a few detected DM events are required.  Neither radio-impurities nor solar neutrinos can mimic the dipole signature expected from dark matter~\cite{OHare:2015utx}.  Its magnitude and experimental robustness makes the directional dipole signature a ``smoking gun" signature of DM.

\begin{figure}[ht!]
\begin{center}
\includegraphics[width=0.9\columnwidth]{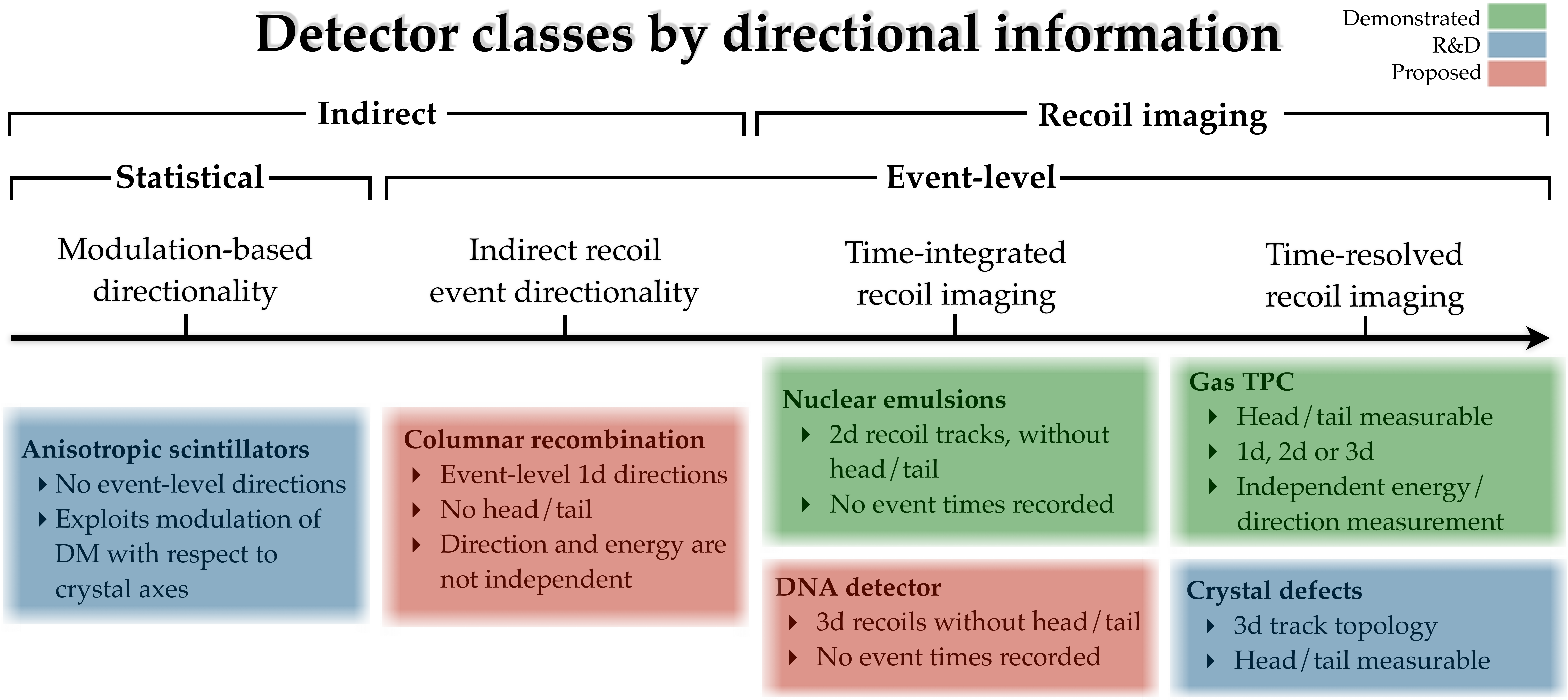}
\caption{Various classes of directional detector ordered roughly (from left to right) from least to most directional information that they are able to access. The minimal amount of directionality is for an experiment that can only infer some kind of direction-dependence from a statistical distribution of event information. On the other hand, a fully direction-sensitive experiment is one that can access up to 3D information on each recoil track individually, and in real-time. This is possessed only by gas TPCs and crystal defect approaches. Figure adapted from Ref.~\cite{Vahsen:2021gnb}}\label{directional_classification}
\end{center}
\end{figure}

The various strategies for directional recoil detection are reviewed in Ref.~\cite{Vahsen:2021gnb}, and summarized in Fig.~\ref{directional_classification}.  Directional detection can be achieved either directly by imaging the nuclear recoil trajectory, or indirectly by inferring the direction from a proxy variable.  We will see below that low-density gas TPCs are capable of direct recoil imaging. Indirect approaches include anisotropic light yield in crystals, and anisotropic ionization detection due to columnar recombination in liquid or gas, which have not yet been demonstrated at the energies relevant for DM searches.

With a recoil-imaging detector, it is possible to observe the directional galactic dipole signature with as few as 5--10 detected DM events if the performance is at the level of the following or better~\cite{Vahsen:2021gnb}: recoil axis angular resolution $\leq 30^\circ$; efficiency for correctly detecting the recoil head/tail $>$ 80\%; offline rejection of electron-recoil background by factors $>10^5$.

The NEWS collaboration (based in Japan and Italy) is pursuing directional recoil detection in emulsions~\cite{NEWS:2016fyf} and making good progress.  Outside the particle physics community, quantum defects in wide-bandgap semiconductors have also been proposed to achieve directional sensitivity at solid-state densities \cite{Rajendran:2017ynw,Marshall:2020azl,Ebadi:2022}.  The incident particle's direction would be stored as a durable, submicron track of crystal lattice damage, which could be mapped using solid-state quantum sensing methods. Early R\&D has focused on nitrogen-vacancy defects in diamond \cite{Marshall:2020azl,Marshall:2021kjk,Marshall:2021xiu} to establish the feasibility of directional readout of such tracks. 
The remaining world-wide groups working on recoil imaging are pursuing low-density gas TPCs, where keV-scale nuclear recoils can be mm in length, while charge diffusion is of order 100~$\upmu$m, which allows for the directional reconstruction of nuclear recoils.

Recently, most groups worldwide that are pursuing directional detection in gas TPCs joined forces to form the CYGNUS Detector R\&D collaboration.  The CYGNUS collaboration proposes to build a $\ge 1000~{\rm m}^3$ recoil-imaging gas target TPC.  The proposed detector consists of a large number of smaller modules, which allows for the large volume to be distributed across multiple underground laboratories. Gas detectors have lower target mass per unit volume than liquid or solid target detectors, but can detect even individual electrons of ionization with $\mathcal{O}(100~\upmu$m) spatial resolution.  As we are rapidly approaching the neutrino fog, DM detectors are guaranteed to see an irreducible background soon. This is a background which can be clearly separated from DM signals via directionality. The neutrino scattering events expected in the neutrino fog can also be exploited as a signal.

The approximate timeline for CYGNUS worldwide is: 1) 2022-2025: 1~m$^3$ detectors to be constructed and start operation in the UK, Japan, Italy, US, and Australia; 2) 2025-2035: 10~m$^3$ detectors: CYGNUS HD10 module (electronic readout) to be jointly constructed and operated in the US; a CYGNO detector (optical readout) program is planned and funded in Italy; detectors in Japan, the UK, and Australia are also planned; and 3) 2024-2042+: 1000~m$^3$ detectors, with construction of the facility to begin in 2030.

The largest directional DM detector prototypes to date have been 1~m$^3$ in volume, and were built by the DRIFT~\cite{Battat:2016xxe} and DMTPC~\cite{Ahlen:2010ub} collaborations. Both detectors were designed to search for 100-GeV DM particles, and have limited directionality for recoil energies below 50~keVr.  Recently, smaller R\&D detectors in the US have shown that the particle identification and event-level recoil directionality required for a directional discovery with only 5-10 events can be achieved even at sub-10-keV energies. Modern MPGD-based detectors utilizing electronic readout~\cite{Jaegle:2019jpx} with charge multiplication gains exceeding 3000--9000, result in ionization threshold of order 30~eV, and detection of sub-10-keV recoils. CYGNUS HD recently achieved both the desired low-energy particle identification and directional capabilities with 3D convolutional neural networks (3D CNNs). The desired end result would be a CYGNUS detector operating at the fundamental performance limit where individual primary electrons are counted in 3D at $100~\upmu$m$^3$ spatial resolution, with diffusion kept minimal even at large drift lengths through negative ion drift. Recent R\&D with GridPix charge readout ~\cite{Ligtenberg:2021viw} has demonstrated the feasibility of this on a smaller scale. The CYGNUS-HD members in the US are currently building prototypes to demonstrate sub-10-keV directionality at the 40~L and then 1000~L module scale.  For further detail, we refer the reader to Refs.~\cite{Vahsen:2021gnb,Vahsen:2020pzb,SnowmassIF53}.

\appendix
\newpage 
\section{Table of recent, active, and planned experiments}
%
%
%

\begin{table}[!ht]
    \centering
    \tiny
    \begin{tabular}{|l|l|l|l|l|l|l|l|}
    \hline
        Name & Detector & Target & Active Mass & Location of Experiment & Status & Start\_Ops & End\_Ops \\ \hline
        ~ & ~ & ~ & ~ & ~ & ~ & ~ & ~ \\ \hline
        XMASS & Scintillator & LXe & 832 kg & Kamioke & Ended & 2010 & 2019 \\ \hline
        XENON10 & TPC & LXe & 62 kg & LNGS & Ended & 2006 & 2008 \\ \hline
        XENON100 & TPC & LXe & 62 kg & LNGS & Ended & 2012 & 2016 \\ \hline
        XENON1T & TPC & LXe & "1,995 kg" & LNGS & Ended & 2017 & 2019 \\ \hline
        XENON1T (Ionization) & TPC Ioniz.-only & LXe & "1,995 kg" & LNGS & Ended & 2017 & 2019 \\ \hline
        XENONnT & TPC & LXe & "7,000 kg" & LNGS & Construction/Run & 2021 & 2025 \\ \hline
        LUX & TPC & LXe & 250 kg & SURF & Ended & 2013 & 2016 \\ \hline
        LUX (Ionization) & TPC Ioniz.-only & LXe & 250 kg & SURF & Ended & 2017 & 2019 \\ \hline
        LZ & TPC & LXe & "8,000 kg" & SURF & Construction/Run & 2021 & 2025 \\ \hline
        PandaX-II & TPC & LXe & 580 kg & CJPL & Ended & 2016 & 2018 \\ \hline
        PandaX-4T & TPC & LXe & "4,000 kg" & CJPL & Running & 2021 & 2025 \\ \hline
        LZ HydroX & TPC & LXe+H2 & "8,000 kg" & SURF & R\&D & 2026 & ~ \\ \hline
        Darwin / US G3 & TPC & LXe & "50,000 kg" & LNGS/SURF/Boulby & Planning & 2028 & 2033 \\ \hline
        ~ & ~ & ~ & ~ & ~ & ~ & ~ & ~ \\ \hline
        DEAP-1 & Scintilator & LAr & ~ & ~ & Ended & 2007 & 2011 \\ \hline
        DEAP-3600 & Scintillator & LAr & "3,300 kg" & SNOLAB & Running & 2016 & 202X \\ \hline
        DarkSide-50 & TPC & LAr & 46 kg & LNGS & Ended & 2013 & 2019 \\ \hline
        Darkside-LM (Ionization) & TPC Ioniz.-only & LAr & 46 kg & LNGS & Ended & 2018 & 2019 \\ \hline
        Darkside-20k & TPC & LAr & 30 t & LNGS & Planning/Construct & 2025 & 2030 \\ \hline
        ARGO & TPC or Scintillator & LAr & 300 t & SNOLAB & Planning & 2030 & 2035 \\ \hline
        GADMC & TPC & LAr & ~ & ~ & Planning & 2030 & ~ \\ \hline
        ~ & ~ & ~ & ~ & ~ & ~ & ~ & ~ \\ \hline
        DAMA/LIBRA & Scintillator & NaI & 250 kg & LNGS & Running & 2003 & ~ \\ \hline
        ANAIS-112 & Scintillator & NaI & 112 kg & Canfranc & Running & 2017 & 2022 \\ \hline
        COSINE-100 & Scintillator & NaI & 106 kg & YangYang & Running & 2016 & 2021 \\ \hline
        COSINE-200 & Scintillator & NaI & 200 kg & YangYang & Construction & 2022 & 2025 \\ \hline
        COSINE-200 South Pole & Scintillator & NaI & 200 kg & South Pole & Planning & 2023 & ? \\ \hline
        COSINUS & Bolometer Scintillator & NaI & ? & LNGS & Planning & 2023 & ? \\ \hline
        SABRE PoP & Scintillator & NaI & 5 kg & LNGS & Construction & 2021 & 2022 \\ \hline
        SABRE (North) & Scintillator & NaI & 50 kg & LNGS & Planning & 2022 & 2027 \\ \hline
        SABRE (South) & Scintillator & NaI & 50 kg & SUPL & Planning & 2022 & 2027 \\ \hline
        ~ & ~ & ~ & ~ & ~ & ~ & ~ & ~ \\ \hline
        CDEX-10 & Ionization (77K) & Ge & 10 kg & CJPL & Running & 2016 & ? \\ \hline
        CDEX-100 / 1T & Ionization (77K) & Ge & 100-1000 kg & CJPL & Planning & 202X & ~ \\ \hline
        ~ & ~ & ~ & ~ & ~ & ~ & ~ & ~ \\ \hline
        SuperCDMS & Cryo Ionization & Ge & 9 kg & Soudan & Ended & 2011 & 2015 \\ \hline
        CDMSLite (High Field) & Cryo Ionization & Ge & 1.4 kg & Soudan & Ended & 2012 & 2015 \\ \hline
        CDMSLite (High Field) & Cryo Ionization & Ge & 1.4 kg & Soudan & Ended & 2012 & 2015 \\ \hline
        CDMS-HVeV Si & Cryo Ionization HV & Si & 0.9 g & Surface Lab & Ended & 2018 & 2018 \\ \hline
        SuperCDMS CUTE & Cryo Ionization / HV & Ge/Si & 5 kg/1 kg & SNOLAB & Running & 2020 & 2022 \\ \hline
        SuperCDMS SNOLAB & Cryo Ionization / HV & Ge/Si & 11 kg/3 kg & SNOLAB & Construction & 2023 & 2028 \\ \hline
        EDELWEISS III & Cryo Ionization & Ge & 20 kg & LSM & Ended & 2015 & 2018 \\ \hline
        EDELWEISS III (High Field) & Cryo Ionization HV & Ge & 33 g & LSM & Running & 2019 & ~ \\ \hline
        CRESST-II & Bolometer Scintillation & CaWO4 & 5 kg & LNGS & Ended & 2012 & 2015 \\ \hline
        CRESST-III & Bolometer Scintillation & CaWO4 & 240 g & LNGS & Ended & 2016 & 2018 \\ \hline
        CRESST-III (HW Tests) & Bolometer Scintillation & CaWO4 & ~ & LNGS & Running & 2020 & ~ \\ \hline
        ~ & ~ & ~ & ~ & ~ & ~ & ~ & ~ \\ \hline
        COUPP & Bubble Chamber & CF3I & 4 kg & SNOLAB / Fermilab & Ended & 2011 & 2012 \\ \hline
        PICASSO & Superheated Droplet & C4F10 & 3 kg & SNOLAB & Ended & ~ & 2017 \\ \hline
        PICO-2 & Bubble Chamber & C3F8 & 2 kg & SNOLAB & Ended & 2013 & 2015 \\ \hline
        PICO-40 & Bubble Chamber & C3F8 & 35 kg & SNOLAB & Running & 2020 & ~ \\ \hline
        PICO-60 & Bubble Chamber & "CF3I,C3F8" & 52 kg & SNOLAB & Ended & 2013 & 2017 \\ \hline
        PICO-500 & Bubble Chamber & C3F8 & 430 kg & SNOLAB & Construction/Run & 2021 & ~ \\ \hline
        ~ & ~ & ~ & ~ & ~ & ~ & ~ & ~ \\ \hline
        DRIFT-II & Gas Directional & CF$_4$ & 0.14 kg & Boulby & Ended & ~ & ~ \\ \hline
        NEWAGE-03b' & Gas Directional & CF4 & 14 g & Kamioka ~ & Running & 2013 &2023 \\ \hline
        MIMAC & Gas Directional & CF$_4$+CHF$_3$+C$_4$H$_{10}$ & ~ & LSM (Modane) & Running & 2012& \\ \hline
        CYGNO & Gas Directional & He + CF$_4$ & 0.5 - 1~kg & LNGS & Planning & 2024 & ~ \\ \hline
        CYGNUS & Gas Directional & He + SF$_6$/CF$_4$ & ~ & Multiple sites & Planning & ~ & ~ \\ \hline
        NEWS-G & Gas Drift & CH4 & ~ & LSM & Ended & 2017 & 2019 \\ \hline
        NEWS-G & Gas Drift & CH4 & ~ & SNOLAB & Construction/Run & 2020 & 2025 \\ \hline
        ~ & ~ & ~ & ~ & ~ & ~ & ~ & ~ \\ \hline
        DAMIC & CCD & Si & 2.9 g & SNOLAB & Ended & 2015 & 2015 \\ \hline
        DAMIC & CCD & Si & 40 g Si & SNOLAB & Ended & 2017 & 2019 \\ \hline
        DAMIC100 & CCD & Si & 100 g Si & SNOLAB & Not Built & ~ & ~ \\ \hline
        DAMIC-M & CCD Skipper & Si & 1 kg Si & LSM & Construction/Run & 2021 & 2024 \\ \hline
        SENSEI & CCD Skipper & Si & 2 g Si & Fermilab u/g & Running & 2019 & 2020 \\ \hline
        SENSEI & CCD Skipper & Si & 100 g Si & SNOLAB & Construction/Run & 2021 & 2023 \\ \hline
        Oscura & CCD Skipper & Si & 10 kg Si & SNOLAB & Planning & 2024 & 2028 \\ \hline
        SNOWBALL & Supercooled Liquid & H2O & ~ & ~ & Planning & ~ & ~ \\ \hline
        ~ & ~ & ~ & ~ & ~ & ~ & ~ & ~ \\ \hline
        ALETHEIA & TPC & He & ~ & China Inst. At. Energy & R\&D & ~ & ~ \\ \hline
        TESSERACT & Cryo TES & He & ~ & LBNL & R\&D & ~ & ~ \\ \hline
    \end{tabular}
\end{table}


\newpage  
\printbibliography

\end{document}